\documentclass[useAMS,usenatbib]{mn2e}

\usepackage{graphicx} 
\usepackage{epstopdf}

\newcommand       \be           {\begin{equation}}
\newcommand       \ee           {\end{equation}}
\newcommand       \bea          {\begin{eqnarray}}
\newcommand       \eea          {\end{eqnarray}}
\newcommand       \apj          {ApJ}
\newcommand       \apjl         {ApJL}
\newcommand       \aap          {A\&A}

\newcommand       \mnras        {MNRAS}
\newcommand       \araa         {ARA\&A}
\newcommand       \aj           {AJ}
\def\simlt{\mathrel{\hbox{\rlap{\hbox{\lower4pt\hbox{$\sim$}}}\hbox{$<$}}}}
\def\simgt{\mathrel{\hbox{\rlap{\hbox{\lower4pt\hbox{$\sim$}}}\hbox{$>$}}}}
\newcommand{\gtsim}{\lower.8ex\hbox{$\; \buildrel > \over \sim \;$}}
\newcommand{\lessim}{\lower.8ex\hbox{$\; \buildrel <\over \sim \;$}}

\title[Modeling the spectral evolution of PWNe inside SNRs]{Modeling spectral evolution of PWNe inside SNRs}
\author[N. Bucciantini, at al.]{N. Bucciantini$^{1,2}$\thanks{E-mail:
    nbucciantini@nordita.org}, J. Arons$^{2,3}$, E. Amato$^{4}$\\
$^{1}$NORDITA, Roslagstullsbacken 23, 106 91 Stockholm, Sweden\\
$^{2}$Astronomy Department and Theoretical Astrophysics Center,
  University of California, Berkeley, 601 Campbell Hall,\\ Berkeley CA,
  94720\\ 
$^{3}$Department of Physics, University of California,
  Berkeley, Le Conte Hall, Berkeley, CA 94720\\ 
$^{4}$ INAF, Osservatorio di Arcetri, Firenze, L.go Fermi 5, 50125, Firenze, Italy}
\begin{document}

\date{Accepted . Received ; in original form }

\pagerange{\pageref{firstpage}--\pageref{lastpage}} \pubyear{????}

\maketitle

\label{firstpage}

\begin{abstract}

We present a new model for the spectral evolution of Pulsar Wind Nebulae inside 
Supernova Remnants. The model couples the long-term dynamics of these systems, 
as derived in the 1-D approximation, with a 1-zone description of the spectral evolution 
of the emitting plasma. Our goal is to provide a simplified theoretical description that can
be used as a tool to put constraints on unknown properties of PWN-SNR systems: a 
piece of work that is preliminary to any more accurate and sophisticated modeling. 
In the present paper we apply the newly developed model to a few objects of different 
ages and luminosities. We find that an injection spectrum in the form of a broken-power 
law gives a satisfactory description of the emission for all the systems we consider. 
More surprisingly, we also find that the intrinsic spectral break turns out to be at a similar 
energy for all sources, in spite of the differences mentioned above. We discuss the implications
of our findings on the workings of pulsar magnetospheres, pair multiplicity and on the particle 
acceleration mechanism(s) that might be at work at the pulsar wind termination shock.
\end{abstract}

\begin{keywords}
supernova remnants; pulsars: general; stars: winds, outflows; radiation mechanisms: non-thermal
\end{keywords}

\section{Introduction}
\label{sec:int}
Pulsars (PSRs) are known to release most of their rotational energy in the form of a 
relativistic magnetized wind, whose particle component is mostly made
of electron-positron pairs, with, possibly, a minority of ions, while the magnetic field
is almost purely toroidal far away from the light cylinder (e.g. \cite{goldreich69}).
The wind is initially confined by the slowly expanding Supernova Remnant (SNR),
which leads to the formation of a termination shock at some distance from the pulsar,
and of a bubble of relativistically hot fluid beyond it. This is the so-called 
Pulsar Wind Nebula (PWN), that shines through synchrotron and Inverse-Compton 
emission from radio wavelengths up to $\gamma$-rays.

The typical energy that a pulsar injects into the PWN during its entire
lifetime is about $10^{-2}$ of the typical energy of the SN explosion
($10^{51}$ erg). Therefore, the presence of an energetic PSR, has little effect
on the global evolution of the SNR. In contrast, the evolution of
the PWN strongly depends on its interaction with the SNR (see
Sec.~\ref{sec:neb}).

Modeling of the evolution and dynamical properties of PWN-SNR systems
has much improved since the pioneering one-dimensional
analytical works by \citet{rees74} and \citet{kennel84a}. In recent years, 
the impressive advances in X-ray observations, due to {\it CHANDRA} 
and {\it XMM-Newton}, have led to a renewed interest in these objects.
A major effort has been devoted to explaining the observed X-ray properties 
of the Crab Nebula and other young systems through multi-dimensional 
time-dependent approaches to their dynamics \citep{komiss04,ldz04,ldz05}.
Several authors \citep{blo01,swa01,me04b,dejager08} have
presented numerical evolutionary models, extended to advanced ages
(typically up to $\sim 10^5$ yr), and to the case of fast moving pulsars
\citep{swa04}, but in fact, fully multidimensional models are still lacking
in these cases. The reason for this is that detailed modeling of older systems
is far more difficult than for the young ones: the latter are brighter, hence 
providing us with high quality multi-frequency data, and are often better 
constrained in terms of the most relevant physical quantities, such as age, 
distance and pulsar energetics.  

The overall evolution of old objects is much more complex than the smooth
expansion appropriate to describe young systems. In general, even within 
the most simplistic approach, the long-term evolution of a PWN-SNR system
depends on many parameters: the SN energy, the mass of the ejecta, their
density structure, the density of the ISM, the PSR luminosity, its spin-down 
history \citep{me04b}. A model for particle injection and spectral evolution 
must be added to all of this, if the aim is that of deriving information from
direct comparison with observations. At present, multi-dimensional numerical 
simulations are not suitable to properly investigate the large parameter space 
that older systems may span. 

The situation is actually delicate also for many young systems.
Indeed, accurate multidimensional results, including emission maps
to be directly compared with observations, have been so far presented 
only for the Crab Nebula. But this is a very special object, for which measurements 
are available through almost the entire electromagnetic spectrum and
almost all the aforementioned parameters are known, including the age 
and the pulsar rotational history (initial spin frequency and braking index).
For other young systems, like 3C58 or MSH 15-52, the quantity 
and quality of information available is much poorer than in the case of Crab and
simplified evolutionary models are again the only possible tool for 
investigation, at least for a start.

Models for the evolution of the emission properties of PWNe have a 
long history, from the work by \citet{pacini73}, to the most recent 
developments by \citet{dejager08}, and \citet{gelfand09}. 
These later works have attempted at taking into account both the 
dynamics of PWN-SNR systems, as derived from accurate numerical 
simulations, and the spectral evolution of the particle distribution
function. In particular, the work by \citet{gelfand09} is exemplary for
showing the richness of behaviors that is found even within the 
one-zone treatment, and the degree of complexity that one can investigate 
using simple tools.

While constraining the unknown parameters is preliminary to detailed 
multi-dimensional modeling of the systems, which can only sample a 
much more limited range of parameters, the question arises of how 
strong and reliable the results obtained with a 1-zone approach are. 
The recent success of 2-d models of PWNe at reproducing so many 
of the observed features has interesting consequences in this respect,
in that it lends support to the 1-zone description over the 1-dimensional 
approximation.
The latter  inevitably leads to onion-like structures \citep{kennel84b}, 
where particles injected at earlier times are located at larger radii, with a 
one-to-one correspondence between the properties of the particle distribution 
function and the distance from the central PSR. Comparison of multidimensional 
models with observations has shown that such ordered structure is not what is 
realized in PWNe, where an overall turbulent flow is injected at the termination 
shock and survives through most of the nebular extent \citep{camu09}. For the typical flow times, 
that are much smaller than the PWN age, this turbulent flow-structure leads to 
efficient mixing, which is what a 1-zone description is based on. This also
finds some direct hints from observations where detailed data are available, as 
in the case of the Crab Nebula:
the remarkable homogeneity of radio spectral index across the nebula, and the 
extent of the X-ray emitting region which exceeds the predictions of any simple
1-dimensional flow model in the absence of particle diffusion far more efficient 
than the Bohm rate \citep{amato00}.

In the following we present a simplified evolutionary model, similar in
spirit to the work by  \citet{gelfand09}. Our approach combines analytical 
results and numerical simulations for the description of the dynamics of the
PWN-SNR system and the evolution of the particle distribution function within 
the PWN: the key features of the interaction between a PWN and the 
surrounding SNR are adequately reproduced during different evolutionary stages.
The model is here applied to a few systems of various ages, both young and 
older ones with the aim of showing how this kind of modeling can help
the extraction of 
information on the system properties from multi-wavelength observations: 
we try to clarify how these properties depend on the different parameters, also 
discussing potential degeneracies and the observations that would best serve to 
disentangle them. 

Special attention is devoted to the particularly interesting and difficult task 
of constraining the particle injection in PWNe. The relevance of this issue in
terms of pulsar physics and shock acceleration physics deserves a somewhat
extended discussion, which will be the topic of next section (Sec.~\ref{sec:mul}).

The rest of the paper is organized as follows. Sec.~\ref{sec:neb}, 
Sec.~\ref{sec:par} and Sec.~\ref{sec:emis} describe, respectively, how the 
nebular evolution is implemented in our model, the scheme we adopt to 
follow the evolution of the particle distribution function and the emission model.
In Sec.~\ref{sec:res} we show the application of our model to several objects, and,
finally, in Sec.~\ref{sec:dis} we discuss the implications of our findings.

\section{Particle injection in PWNe}
\label{sec:mul}

Forty years of research on Rotation Powered Pulsars have led to a
conundrum:  observations of young Pulsar Wind Nebulae (PWNe) have
clearly established large particle injection rates into the nebulae,
well in excess of the electrodynamic minimum suggested by
\citet{goldreich69}. The latter corresponds to: 

\begin{eqnarray}
\dot{N}_{R} & = & \frac{c\Phi}{e} = \left( \frac{ c I \Omega \dot{\Omega} }{e^2} \right)^{1/2} \nonumber \\
&  = & 7.6 \times 10^{33} \left( \frac{I_{45}}{P_{33}^3}  \frac{\dot{P}}{4 \times 10^{-13}} \right)^{1/2} 
   \; {\rm s}^{-1}\ ,
\label{eq:RotRate}
\end{eqnarray}
where $\Phi = \sqrt{L /c} $ is the magnetospheric potential, $L = I \Omega
\dot{\Omega} $ is the spin down luminosity, with $I$ the neutron star's
moment of inertia ($I_{45}=I/10^{45}$ g cm$^2$), $\Omega = 2\pi /P$ the 
stellar angular velocity and $P$ the rotation period ($P_{33} = P/33$ msec).
The simplest estimates \citep{rees74, kennel84a} show that X-ray emitting 
PWNe \citep{kargaltsev08} with synchrotron cooling times of the X-ray
emitting particles well less than the nebular ages have electron (and
positron) injection rates much larger than shown in Eq.~(\ref{eq:RotRate}):
values reported in the literature for the multiplicity $\kappa =
\dot{N}/\dot{N}_R$ are typically $\sim 10^4$, based on  analysis of
the X-ray emission, a result consistent with theoretically derived
pair creation rates for the young, high voltage pulsars which have
been subjected to such analysis [{\it e.g.} \citet{kennel84a, gae02,
  hibschman01}]. Indeed, the fact that the inferred injection rates
are this large is one of the major pillars of support for the
theoretical conclusion that pulsars have substantial outflows of
$e^\pm$, this being the only known means through which such cool
objects can have winds denser than the electrodynamic minimum.

Such high density flows with $\kappa \gg 1$ support the use of the
force-free limit of MHD theory in modeling the torques on rotation
powered neutron stars \citep{kala09, spitkovsky06}. As already 
mentioned, MHD theory has been quite successful in modeling the 
multidimensional dynamics and appearance of young PWNe 
\citep{komiss04, ldz04, ldz05,vol08}. Such models confirm the early 
inference \citep{rees74,kennel84a} that the system behaves as if 
just upstream of the pulsar wind termination shock (TS) the plasma 
has low magnetization $\sigma_w \equiv (B^2/8\pi
\gamma_w n_\pm m_\pm c^2) \ll 1$ ($n_\pm$ is the number density of
pairs, the particle density is $2n_\pm$): typically, the average of
$\sigma_w$ in latitude with respect to the rotation axis
is\footnote{This analysis has been done in full only for the Crab
Nebula, however.} $\sim 0.02$. 

The nebular dynamics in the MHD model is insensitive to the specific
value of the upstream 4-velocity $u_w =c\beta_w \gamma_w$, as long as
$\gamma_w \gg 1$.  By applying their 1D MHD model to the optical and
harder photon emission in the Crab, \cite{kennel84b} inferred
$\gamma_w \approx 10^{6.5}$, with $\sigma_w \approx 0.005$. More
modern models (\citealt{ldz05, vol08}) require latitude averaged
$\sigma_w$ to be somewhat larger (so that magnetic hoop stress can
create the jet component of the torus-jet structure), while account of the
high energy synchrotron emission requires particle spectra at the TS
with parameters similar to the 1D Kennel and Coroniti model: again 
$\gamma_w \sim 10^6$, although this is not explicitly stated 
since the distribution functions were not tied to the specifics of the MHD flow.

If the apparent low value of $\sigma_w$ means that the wind just upstream 
of the TS really is weakly
magnetized, $\sigma_w$ and $\gamma_w$ are closely tied to the pair
multiplicity $\kappa_\pm = \kappa/2$, since when $\sigma \ll 1$ the
wind carries the rotational energy lost from the pulsar in kinetic
energy of the flow, $L = \dot{M} c^2 \gamma_w$.  Since
$\dot{M} = 2 \kappa_\pm m_\pm \dot{N}_R = 2 \kappa_\pm m_\pm c\Phi
/e$, and $L = c\Phi^2$, then $\gamma_w = e\Phi /2 m_\pm c^2
\kappa_\pm$.  In the case of the Crab Nebula and pulsar, $\Phi = 4
\times 10^{16}$ V, while extant theoretical models of pair creation,
from polar caps, slot gaps or outer gaps (\citealt{hibschman01,
  cheng07, hirotani08, harding04}) all yield $\kappa_\pm \sim 10^4$,
for young pulsars with $\Phi > 10^{15}$V. Thus theory also says
$\gamma_w \sim 10^6$ - if one confines one's analysis to the high
frequency emission from PWNe, theory and observation appear to be in
good accord\footnote{We neglect a possible component of heavy ions in
the wind's composition(\citealt{gallant94, spitkovsky04}): this subject
will be discussed later on}.

Using the X-ray emitting particles in PWNe, whose synchrotron cooling
times are short, takes advantage of such PWNe being calorimeters for
the contemporary particle injection rates, which gives the results
some independence from the uncertainties of evolutionary
models. However, radio and infrared emitting particles are much more
numerous than the X-ray emitting ones, due to the rapid decline of 
synchrotron emissivity with declining particle energy (\citealt{kennel84b, gallant02}). 
Models of the underlying pulsar must be able to account for {\it all} the radiating
relativistic particles found in the PWNe - the spectral continuity in
several systems demonstrated in the results described below suggests
the lower energy particles are indeed injected by the pulsar, rather
than being accelerated out of the non-relativistic material often
found embedded within PWNe. Thus measuring the full relativistic
particle content in PWNe is an important experimental input into
modeling of pulsars, distinct from the modeling of the energy input into the
nebulae. For the latter the force free model (\citealt{spitkovsky06, kala09}) 
provides the essential description, but unfortunately this is independent 
of the multiplicity, so long as $\kappa \gg 1$. 

We study several young PWNe (Crab, 3C58, B1509, Kes75) for which the
data can be used to reasonably constrain the pair injection rate - in
principle one would like to use nebulae with ages known, and reasonably
complete (including near and far infrared and millimeter) spectral
energy distributions (SEDs) are
required. We show that all these systems have spectral continuity from
the radio through the infrared to the X-ray bands, suggesting a single
source for the radiating particles is present in each system.  We use
these full SEDs to derive new estimates for the pair creation
multiplicity.  We discuss the constraints these results set on PSR
pair production gap models - Polar Caps, Slot Gaps, and Outer Gaps - 
concluding that no existing model adequately explains particle
injection rates. 

We also discuss alternate hypotheses, that low energy
particles are picked up from thermal gas in the nebulae, or are
fossils left over from some unnamed acceleration process in the early
history of the nebula (\citealt{atoyan96}), or represent the
effects of a second acceleration mechanism operating at low energy (\citealt{gallant02}),
such as cyclotron acceleration stimulated by a heavy ion component of
the wind (\citealt{hoshino92,amato06}), or more generally any component with
Lorentz factor greater than $\gamma_w$. We use the spectral continuity to argue 
that such a two component model is unlikely, implying some additional 
piece of physics needs to be added to the shock acceleration model to 
account for these broken power-laws, in addition to the extra physics 
needed to account for the total injection rate.

\section{Nebular evolution}
\label{sec:neb}

There are three phases of interest in the evolution of a PWN. In this section
we briefly describe them and explain how we model each of them.

We consider objects of relatively young age (less than 40,000 yr) 
with the PSR still inside the SNR, and assume spherical symmetry,
neglecting the pulsar proper motion (see \citet{swa04} for a discussion of this
point). We also assume, in deriving the PWN evolution, that
radiation losses of the particles are negligible. Given the weak
dependence of the PWN radius on the PSR luminosity, and the fact that
radiation losses only affect the particle energy content and not the
magnetic energy content, we do expect this to be a good
approximation. This allows us to use analytical formulae, and
substantially reduce the computational requirements.

\subsection{SNR evolution}

In order to model the evolution of a PWN inside a SNR we first need to
solve for the evolution of the SNR itself. An excellent semi-analytic
model for the evolution of an SNR without an embedded PWN was 
provided by \citet{tru99}.
Using pressure balance between the shocked ISM downstream of
the outer forward shock, and the shocked ejecta beyond the 
reverse shock, they solve for the evolution of both the outer forward
shock and the inner reverse shock from the early ejecta dominated
phase \citep{ham84} to the later Sedov phase \citep{ost88}. It is shown that
the evolution can be cast in dimensionless form by using the following
characteristic variables:
\begin{eqnarray}
R_{ch}=M_{ej}^{1/3}\rho_o^{-1/3}\\
t_{ch}=E_{SN}^{1/2}M_{ej}^{5/6}\rho_o^{-1/3}
\end {eqnarray}
where $\rho_o$ is the density of the ISM (assumed to be uniform),
$E_{SN}$ is the kinetic energy of the supernova explosion, whose
canonical value is $10^{51}$ erg,
and $M_{ej}$ is the mass of the ejecta. The only free parameter in
the Truelove \& McKee model is the mass distribution in the ejecta. 
Unfortunately, this is not directly constrained by observations, and choices in the
literature are mostly based on theoretical assumptions: for example \citet{che82}
suggested a distribution with an inner plateau and outer steep
profile; while \citet{tru99} focus on the case of self-similar ejecta with
a density distribution $\rho_{ej}\propto r^{-\alpha} t^{\alpha-3}$ and a velocity
profile $v_{ej}\propto r$. Many recent numerical works on young PWNe
have indeed adopted the latter distribution, with great success in
reproducing the observations \citep{swa01,swa03,me03,me04,ldz04,ldz05}.

Once the evolution of the SNR is known, it can be used to constrain
the evolution of the PWN.

\subsection{Free expansion phase}

The first phase of the PWN evolution is generally referred to as
{\it free-expansion phase}, and has the PWN expanding inside
the cold ejecta. This phase lasts for a few thousand years, until the
PWN reaches the reverse shock. In this phase, the evolution of the PWN 
is independent of the evolution of the SNR shell, because no contact has
been established between the two yet. In the case of self-similar
ejecta and constant PSR luminosity $L$ a solution for the radius of the PWN
as a function of time has been known for a long time \citep{che92,swa01}:
\begin{equation}
R(t)\simeq L^{\frac{1}{5-\alpha}}E_{SN}^{\frac{3+\alpha}{10-2\alpha}}M_{ej}^{-1/2}t^{\frac{6-\alpha}{5-\alpha}}\ .
\end{equation}
In the more general case in which the PSR luminosity changes in time
according to
\begin{equation}
L(t)=L_o/(1+t/\tau)^\beta\ ,
\end{equation}
\citet{me04} have shown that it is possible to find an analytic
solution and they provide a series expansion for it, showing that 
already the first few leading terms of the series provide an excellent 
approximation. In this work we adopt this approach, that allows us 
to properly include in our model the spin-down properties of the PSR.

The free expansion phase lasts as long as the radius of the PWN is
smaller than the radius of the reverse shock computed using the
Truelove \& McKee model. Once the PWN reaches the reverse shock, 
the reverberation phase begins.

\subsection{Reverberation phase}
Once the PWN has reached the SNR reverse shock, it is 
in contact with the shock heated ejecta. The pressure of the hot ejecta 
is higher than the pressure of the relativistic material inside the PWN. As a
consequence, the PWN expansion halts, and the system starts
contracting \citep{swa01}. 

An analytic model for this phase is not available,
however the evolution can be treated within the so called thin-shell 
approximation \citep{giu82}: the evolution of the PWN outer radius is described
in terms of the evolution of a thin shell of material enclosing a mass $M_{sw}$ 
equal to the mass of ejecta that has been swept up by the PWN up to
the beginning of the reverberation phase. This shell is bounded on the inner 
side by a hot relativistic plasma with pressure $P_{in}$, and on the outer side 
by the shock heated ejecta with pressure $P_{out}$:
\begin{equation}
M_{sw}{\ddot R(t)}=4\pi R(t)^2 (P_{in}(t)-P_{out}(t))\ .
\label{eq:Rsw}
\end {equation} 
The value of the PWN pressure is computed numerically using energy
conservation:
\begin{equation}
P_{in}(t)=P_{in}(t_r)+\frac{1}{4\pi R(t)^4}\int_{t_r}^{t}L(t)R(t)dt\ ,
\label{eq:pin}
\end{equation}
where $t_r$ is the time at which the reverberation phase begins.
It is less clear what $P_{out}$ should be.
A lower limit is given by the pressure that the shock heated ejecta
would have in the absence of a PWN. An upper limit corresponds
to the pressure of the Sedov solution.
Numerical simulations show that the interaction of the SNR with the 
PWN leads to additional heating of the ejecta, due to sound waves 
that are launched inside the SNR during the reverberation phase 
(see Fig.~3 of \citet{me03}). On the other hand, it takes longer for the SNR
to relax to the Sedov solution, while during the first compression the value of 
$P_{out}$ is found to be close to about 50\% of the Sedov
value \citep{me03}. In our model we use this fiducial fudge factor.

In 1D simulations \citep{swa01} the reverberation phase is characterized 
by a series of compressions and expansions of the nebula, until the
system relaxes to the Sedov-Taylor phase, which is finally established once the
PWN reaches pressure equilibrium with the SNR. This behavior, which
resembles a damped oscillator, is an artifact of the 1D geometry. 
In more realistic multidimensional regime \citep{blo01} , the evolution of the
PWN during the reverberation phase is subject to strong
Rayleigh-Taylor (RT) instabilities and efficient mixing of the relativistic
material with the SNR matter. From a dynamical point of view this
mixing acts as a viscous term on the evolution of the nebula, and one
might expect, instead of a series of oscillations, an almost complete
relaxation to the Sedov-Taylor solution after the first compression.

Of course in the presence of mixing and the related clumpiness, the 
volume occupied by the relativistic plasma will not be directly related 
to the radial extent of the nebula. However Eq.~\ref{eq:pin} will still 
hold if one interprets $R(t)$ as an effective radius related to the total 
volume of the relativistic plasma, $4\pi R(t)^3/3$, rather than as the radial 
extent of the nebula (which in general might be larger because of
clumpiness). In 1-zone models this effective volume is also all that
matters for computing particle adiabatic losses.

In our model the reverberation phase ends after the first compression,
once the PWN pressure reaches the value of the Sedov-solution for the
SNR.

\subsection{Sedov-Taylor phase} 

Once the pressure inside the PWN reaches the value proper of 
the Sedov solution corresponding to the SNR forward shock, the
Sedov-Taylor phase begins. This usually happens at an age $\sim 10^4$
yr. By this time, because of spin-down, the PSR luminosity can in
general be neglected in the evolution of the system. The speed of the
forward shock, $v_{fs}$, and the post-shock pressure are both given by 
the \citet{tru99} model. The pressure inside the SNR is then assumed to be
equal to the central pressure of the Sedov solution $P_{out}\sim 0.5 \rho_o
v_{fs}^2$.

The nebula evolves according to
\begin{equation}
R(t_s)^4P_{in}(t_s)=R(t)^4P_{in}(t)=R(t)^4P_{out}(t)\ ,
\label{eq:Rst}
\end{equation}
where $t_s$ is the time at which the Sedov-Taylor phase begins and,
again, the radius $R(t)$ is related to the total volume of the
relativistic plasma rather than to the radial extent of the nebula.

One last comment is in order about the role of RT instabilities.
While during the reverberation and Sedov-Taylor phase, re-interpretation
of the radius appearing in Eqs.~\ref{eq:Rsw}, \ref{eq:pin} and \ref{eq:Rst}
in terms of effective volume is necessary, during the initial free-expansion 
phase, RT and mixing are not as important: \citet{me04b} and 
\citet{jun98} have shown that, in general, the radius derived from the 
1D solution, is at most ~ 10-15\% smaller than the true radial extent 
of the nebula. We might then conclude that at least for very young 
objects the 1D radial model provides a reliable estimate both for the 
radial extent of the nebula and for the volume occupied by the relativistic fluid.

\section{Particle Evolution}
\label{sec:par}

Once the evolution of the nebula is known one can compute the
evolution of the particle distribution function. 

The energy of a particle is evolved according to
\begin{equation}
\label{eq:edot}
\frac{dE(t)}{dt}=-\frac{\dot{R}(t)}{R(t)}E(t)-\frac{4\sigma_t}{3m^2c^3}E^2(t) \left(\frac{B(t)^2}{8\pi}+U(t)\right)\ ,
\end{equation}
where $B(t)$ and $U(t)$ are the magnetic field and the
background photon energy density in the nebula, respectively, 
$\sigma_T$ is the Thompson cross section, $m$ the particle mass. 
The magnetic field in the nebula is computed assuming that the 
ratio of magnetic to total energy in the
nebula $\eta_M$ is constant in time ($0<\eta_M<1$), while the
total energy is computed considering the pulsar injection and the
adiabatic losses of the 1D model as described in Sec.~\ref{sec:neb}. The background 
photon energy density includes different contributions: CMB and starlight, 
which are constant in time, and synchrotron (SYN) emission, which is computed,
instead, together with the nebular evolution. We have verified that Inverse 
Compton (IC) losses are in general negligible with respect to SYN losses, 
and SYN-IC is important only in young compact objects, where the magnetic 
field is stronger, and the SYN emissivity higher. 

Given a particle injected at time $t_o$ with energy $E_o$, one
can solve Eq.~\ref{eq:edot} to derive the energy $E(t,t_o,E_o)$ of the particle at
time $t$. An analytic solution of Eq.~\ref{eq:edot}, is not in
general available, neither is its inverse, and the energy evolution has to be 
computed numerically. Once the evolution of the energy is known, then it is
possible to compute the evolution of the particle spectrum according to
\begin{equation}
N(E,t)=\int_E^\infty\dot{N}(E_o,t_o)\frac{\partial t_o}{\partial E}(E,E_o,t)dE_o\ ,
\end{equation} 
where $\dot{N}(E)$ is the particle injection rate per unit energy interval. 

Since no analytic solution is available for Eq.~\ref{eq:edot}, also this equation
must be solved numerically. Given the very short synchrotron lifetime of the
particles at the high energy end of the spectrum we use a Lagrangian scheme 
in energy space. The energy space is originally divided into energy bins whose 
extremes are given by $E_i(t)$. These evolve according to Eq.~\ref{eq:edot}. 
When a bin moves to energies lower than 50 keV, it is removed from the sample. 
New bins are continuously added at the high energy end of the distribution function, 
as the old ones evolve to lower energies. The time step is then dictated by the 
requirement of sufficient energy resolution in the high energy portion of the spectrum.

The total number of particles inside the i-{\it th} bin, at each time, is then computed by
integrating numerically the following equation:
\begin{eqnarray}
N_i(E_{i+1/2}(t),t)=\nonumber\\
N_i(E_{i+1/2}(t_o),t_o)+\int_{t_o}^t dt'\ \int_{E_i(t')}^{E_{i+1}(t')} dE\ \dot{N}(E,t')\ ,
\end{eqnarray}    
where $t_o$ is the initial time at which injection in the bin
begins. This approach guarantees conservation of particle number.

In principle the functional form of $\dot{N}(E,t)$ can be
arbitrarily chosen. 

For the the pairs, we assume a broken
power-law. Given that injection is due to the PSR wind, and that the
only energy scale available in the wind is the PSR voltage, $\epsilon_v=e\Phi$, 
we have decided to use a scale-free model for a broken power-law:
\begin{equation}
\dot{N}(E,t)=C_o(t)
\left\{
\begin{array}{cc}
(E/\epsilon_c)^{-\gamma_1}  & {\rm for}\; \epsilon_m<E<\epsilon_c \\
(E/\epsilon_c)^{-\gamma_2}  & {\rm for}\; \epsilon_c<E<\epsilon_v \\
\end{array}
\right.
\label{eq:partspec}
\end{equation}
where $\gamma_1<2<\gamma_2$, $\epsilon_c$ is the peak of the injected 
energy distribution $E \dot{N}(E,t)$ and $\epsilon_m$ is the
low-energy cutoff, usually found to be small compared to $\epsilon_c$. 
The scale free approximation implies that the ratios $\nu_e=\epsilon_m/\epsilon_v$ 
and $\mu_e=\epsilon_c/\epsilon_v$, are constant in time. $\mu_e$ and $\nu_e$ are 
free parameters of the model, and their value is obtained from a fit to the
data. 

Once $\mu_e$ and $\nu_e$ are fixed, $C_0(t)$ is given by energy conservation. 
We have:
\begin{equation}
\eta_eL(t)=\int_{\epsilon_m}^{\epsilon_v}\dot{N}(E,t)EdE \approx
\frac{\gamma_2 - \gamma_1}{(\gamma_2-2)(2-\gamma_1)}
C_0(t)\epsilon_c^2\\
\end{equation}
with $\eta_e$ the fraction of the PSR luminosity injected into pairs.
The particle number flux in the wind is
\begin{equation}
\dot{N}(t)=\int_{\epsilon_m}^{\epsilon_v}\dot{N}(E,t)dE 
     \approx \frac{C_0(t) \epsilon_c}{\gamma_1 -1}
     \left(\frac{\epsilon_c}{\epsilon_m} \right)^{\gamma_1-1}\ ,
\label{eq:dotN}
\end{equation}
and the average energy/particle in the spectrum (\ref{eq:partspec}) is,
for $\epsilon_m \ll \epsilon_c \ll \epsilon_v$,
\begin{eqnarray}
\langle E \rangle & = & \frac{\eta_e L(t)}{\dot{N}(t)}  = \gamma_w m_e
c^2  \approx \nonumber \\ 
&\approx & \frac{(\gamma_2 - \gamma_1)
  (\gamma_1-1)}{(\gamma_2-2)(2-\gamma_1)} \epsilon_c
\left(\frac{\epsilon_m}{\epsilon_c} \right)^{\gamma_1-1}\ .
\label{eq:avgE}
\end{eqnarray}
In expression \ref{eq:avgE}, we have identified the average energy/pair 
in the spectrum with the upstream flow energy/pair, but we have not 
specifically used the shock jump conditions. This is because, as we will 
discuss in \S \ref{sec:dis}, the distribution (\ref{eq:partspec}) implies 
some additional mechanism for energy redistribution, which may or may not 
be directly connected to the termination shock.
From Eq.~\ref{eq:avgE} it is clear how it is possible to estimate the PSR 
multiplicity through a fit of the model to the PWN emission.

If the wind is characterized by a single value of the Lorentz
factor, then $\gamma_w m_ec^2\simeq \epsilon_c$, $\eta_eL(t)=m_ec^2\gamma_w\dot{N}(t)$.
However, the approximation of a single Lorentz factor might not be
correct, in this case the ratio $\epsilon_m/\epsilon_c$ cannot be
a-priori determined, but is found from requiring a fit to the entire
spectrum. In particular, once an upper limit for $\epsilon_m$ is derived 
from fitting the low frequency radio emission, this can be used to infer
the particle number flux and estimate the average Lorentz factor.
 
We consider the possibility that a minority by number of high energy
particles in the equatorial return current (either positrons, ions or
electrons, depending on return current sign) are part of the PSR wind and
carry a fraction $\eta_p$ of the energy flux. Focusing on the case
when ions are the implied particles, one might look for signature of
their presence. For simplicity we consider them injected with a
mono-energetic spectrum centered at an energy $\gamma_w m_pc^2$
(equivalent to assuming  that they are moving at the same 
Lorentz factor as the pairs injected at $\epsilon_c$).
Ions are affected by different 
loss mechanisms compared to pairs: while radiation losses are negligible, 
p-p scattering and diffusion outside
the nebula might be important. Following the approach of \citet{ama03} we 
assume that both diffusion and $p-p$ scattering remove particles from the
distribution function. Due to these effects the number of ions of energy
$E$ between time $t$ and $t+dt$ changes by:
\begin{equation}
N(E(t+dt),t+dt)=N(E(t),t)e^{-dt(1/\tau_{pp}+1/\tau_{dif})}
\end{equation} 
where $E(t)$ is given by Eq.~\ref{eq:edot}, and where the characteristic $p-p$ timescale is
\begin{equation}
\tau_{pp}\simeq 5(\sigma_on_tc)^{-1}
\label{eq:tpp}
\end{equation}
with $\sigma_o=5\times 10^{-26}$ cm$^2$, and $n_t$ the target proton density.
While the typical diffusion timescale is
\begin{equation}
\tau_{dif}\simeq \frac{R(t)^2}{r_L c}
\end{equation}
where $r_L$ is the ions' Larmor radius.

\section{Emission}
\label{sec:emis}

There are three channels of non-thermal emission in PWNe. From Radio 
frequencies to MeV photon energies, the dominant emission process is synchrotron 
radiation by the accelerated electrons and positrons  gyrating in the nebular 
magnetic field. At higher energies, there are two possible contributions: Inverse 
Compton (IC) emission by the high energy pairs, and gamma-ray emission from 
neutral pion decay, following ``p-p'' scattering between accelerated and target
protons.

The synchrotron luminosity is computed using the ``monochromatic'' approximation. 
The power per unit frequency emitted by a single particle with energy 
$\gamma m_e c^2$ is:
\begin{equation}
S(\nu,\gamma)=\frac{4}{3}\sigma_T c
\frac{B^2}{8\pi}\gamma^2\delta(\nu-\nu_c)
\label{eq:snu}
\end{equation} 
where $\sigma_T$ is the Thompson cross section, $B$ is the nebular
magnetic field, and
\begin{equation}
\nu_c=0.29\frac{3e}{4\pi m c}B\gamma^2
\end{equation}
is the characteristic emission frequency.

Eq.~\ref{eq:snu} has then to be integrated over the pair distribution function (see previous section)
in order to derive the total nebular synchrotron luminosity at a given frequency.

IC is computed using the full Klein-Nishina cross section (\citealt{jones68,blumenthal70})
and assuming scattering over three main target photon fields:
1) CMB, modeled as a prefect black-body, with typical temperature 2.75 K; 
2) synchrotron emission from the PWN, consistently computed as described above; 
3) ``galactic background'', which in principle includes both galactic starlight and any local
contribution in the optical-IR band. This third term is modeled as a suppressed black-body: 
both the temperature (in the range of a few hundreds/thousands Kelvin), and the suppression factor 
(of order $10^{-10}$), are in principle free parameters that can be adjusted. The role of this 
contribution will be discussed in more detail when dealing with specific objects.

Gamma ray emission from neutral pion decay is computed following
\citet{ama03}. In the scaling regime a mono-energetic proton
distribution with energy $E_p$ leads to the following pion injection spectrum
\begin{equation}
\frac{dN_\pi}{dtdE_\pi}=K_\pi E_\pi^{-1}g_\pi(E_\pi/E_p)\ ,
\end{equation}
where
\begin{equation}
g_\pi(x)=(1-x)^{3.5}+e^{-18x}/1.34
\end{equation}
and the constant $K_\pi$ is found through the normalization condition:
\begin{equation}
N(E_p)\left(\frac{dE_p}{dt}\right)_\pi=\int_{m_\pi c^2}^{E_p}\frac{dN_\pi}{dtdE_\pi}E_\pi dE_\pi\ ,
\end{equation}
where $N(E_p)$ is the number of protons with energy $E_p$. 

The number of photons emitted per unit time and energy interval is then
\begin{equation}
J(e_\gamma)=4\left(\frac{dN_{\pi^o}}{dtdE_{\pi^o}}\right)_{E_{\pi^o}=2e_\gamma}.
\end{equation}

The total pion emissivity is then computed integrating over the total
proton energy distribution. The density of target protons $n_t$ that enters 
Eq.~\ref{eq:tpp} is the main unknown: if the protons are well confined within the PWN,
then a reasonable assumption would be to use the swept-up ejecta mass;
if they can leak out then one should consider the total ejecta
mass. A realistic model for proton diffusion in PWN-SNR systems 
is far beyond the scope of this paper: we keep $n_t$ as a free parameter
and discuss the implications of its assumed value whenever relevant. 

\section{Results}
\label{sec:res}

In this section we show the results obtained when applying our model
to a few objects of different ages, both young and old. For each object, 
we will not attempt at obtaining the best possible fit of the data, which in 
our view would not be very significant given the intrinsic limitations of a
one-zone model. In fact, our primary interest is to put some constraints on
the unknown parameters and to clarify which observations would be more relevant 
to improve our understanding of these systems. At the same time we will discuss
the limitations of this approach and make clear which part of our results should be
considered as really reliable.

\begin{table*}
\begin{minipage}{17cm}
\caption{Values of the parameters used in the modeling. \label{parameters}}
\label{tab:1}
\begin{center}
\begin{tabular}{l c c c c c}
\hline
Parameter & Symbol & Crab & 3C58 & B1509-58 & Kes 75 \\
\hline
Supernova explosion energy ($10^{51}$ erg) & $E_{SN}$ & 1 & 1 & 7 & 2.1\\
Mass of the ejecta ($M_\odot$) & $M_{ej}$ & 9.5 & 3.2 & 4.0 & 16.4\\
ISM density (cm$^{-3}$) & $\rho_o$ & 0.1 & 0.01 & 0.001 & 2\\
Ejecta density index & $\alpha $ & 0 & 1 & 1 & 0\\
Initial pulsar luminosity ($10^{38}$erg s$^{-1}$) & $L_o$ & 35 &
0.73 & 49 & 1.66\\
Spin-down time (yr) & $\tau$ & 730 & 3280 & 114 & 226\\
Braking index & $\beta $ & 2.33 & 2 & 2.087 & 2.12\\
Age (yr) & $t$ & 950 & 2100 & 1570 & 650\\
Fraction of $L$ that goes into pairs &
 $\eta_e$ & 0.8 & 0.75 & 0.7 &0.95 \\
Low-energy injection index & $\gamma_1$ & 1.5 & 1.2 & 1.2 & 1.7\\
High-energy injection index & $\gamma_2$ & 2.35 & 2.82 & 2.14 & 2.3\\
Peak energy ratio & $\mu_e$ & 1.54$\times 10^{-5}$ & 6.25$\times
10^{-6}$ & 3.33$\times 10^{-6}$ & $ 10^{-4}$\\
Minimum energy ratio & $\nu_e$ & 1.1$\times 10^{-8}$ & 8.7$\times
10^{-8}$ & 2.8$\times 10^{-8}$ & $2\times 10^{-7}$\\
Break energy (eV) & $\epsilon_c$ & 4$\times 10^{11}$ & 5$\times 10^{10}$ & 1.5$\times 10^{10}$ & 4$\times 10^{11}$\\
Fraction of magnetic energy & $\eta_M$ & 0.11 & 0.5 & 0.53 & 0.005\\
Fraction of $L$ that goes into ions& $\eta_p$ & 0 & 0 & 0 & 0\\
\hline
\end{tabular}
\end{center}
\medskip
\end{minipage}
\end{table*}


\subsection{The Crab Nebula}

The Crab Nebula is the system for which we have the best constraints,
both regarding the pulsar injection properties, and the luminosity at
different wavelengths. Available data extend from low radio
frequencies \citep{bal70,baa72}, through mm/IR \citep{mez86,ban02},
optical/UV  radiation
\citep{ver93,hen92}, X-rays \citep{kui01} and $\gamma$-rays from 
MeV to TeV energies \citep{aha04,alb08,abdo10}. The source distance is estimated 
to be 2 kpc, and the nebular volume is $30$ $pc^3$ \citep{hes08}, corresponding 
to a spherical radius of about 6 $ly$. This is the only system for which we know the
exact age ($950$ years). We also have a reliable estimate of the
braking index (2.51 implying $\beta = 2.33$, \citet{lyn93}), which, when combined with the 
present luminosity $L\simeq 5\times 10^{38}$ erg s$^{-1}$, allows us to derive 
an initial luminosity $L(0)= 3.5\times 10^{39}$ erg s$^{-1}$, and a characteristic 
spin-down timescale $\tau \simeq 730$ years.

Much more uncertain are the properties of the SNR which should
be surrounding the nebula. Several attempts at detecting both the 
forward and the reverse shock have only provided upper limits.
More information can be derived by studying the 
filamentary network surrounding the PWN and comparing the results 
with theoretical expectations based on hydrodynamical models of the 
interaction between the nebula and the surrounding ejecta 
\citep{hes96,jun98,me04b}. The visible remnant contains at least 1-2 
$M_\odot$ of He-rich line-emitting gas and the PWN expands at a speed 
of about $1.5-2\times 10^3$ km s$^{-1}$. Traditional models for the evolution
of the Crab Nebula \citep{che92,swa01,me03,ldz04} have all assumed a flat 
density profile of the ejecta. Steeper ejecta ($\alpha \ge 1$ in Table~\ref{tab:1})
give a mass in the filamentary network higher than what is observed. Assuming 
flat ejecta and a canonical SN energy of $10^{51}$ erg, in order to reproduce the 
proper size of the nebula a mass $M_{ej} = 9.5 M_\odot$ is needed, very close to 
the most recent estimate by \citet{mac08}. The lack of any clearly detected SNR 
shell prevents any reliable estimate of the ISM density and we simply assume 
a fiducial value of $0.1$ cm$^{-3}$.

The PSR and SNR parameters are listed in Table~\ref{tab:1}, together with the
parameters that describe the particle energy distribution at injection. The low energy 
spectral index $\gamma_1$ is fixed according to the radio data, while the high energy index 
$\gamma_2$ is chosen in order to minimize the X-ray residuals. We want to stress 
here that the X-ray emission is concentrated in the central region of the PWN and it is
strongly affected by the details of the flow dynamics just downstream
of the termination shock \citep{ldz04,ldz05}. It is unrealistic to expect
that a simplified one-zone model can provide an accurate description
of the high energy spectrum. A one-zone model can at most provide an
indication of the best power-law index that can fit the data, but more
realistic multidimensional models are necessary to address the emission
properties in the X-ray band.
Two features are interesting to notice: 1) $\gamma_2$ is higher
than the value 2.23 typical of relativistic Fermi shock acceleration
with isotropic scattering in the fluid frame; 
2) there are indications that a single power-law 
cannot reproduce the complete set of X-ray data points. 

Several features are present at high energies, but it is not clear if they 
are intrinsic (a suggestion in this sense has been put forward by \citet{vol08}, 
where they present a multidimensional model of the emission that shows 
similar features), if they are simply due to calibration issues among different 
instruments or even for one instruments at different energies, or if they are due
to temporal variability of the emission properties \citep{me08}.

The present model can reproduce the synchrotron part of the spectrum within
few \% accuracy (greater discrepancies being present at high
energies as discussed above). At present, our particle distribution function
does not include any smooth high-energy cut-off, the reason being the very short
lifetime of high energy particles that prevents efficient numerical
integration of their evolution. However a good match with the
data can be obtained by introducing  {\it a posteriori} an exponential cut-off 
in the particle energy distribution. This reflects in a cut-off in the emission
spectrum at a frequency around $\sim 10^{22}$ Hz.  
Our model requires that 80\% of the pulsar spin-down luminosity goes into 
accelerated pairs. This does not leave much energy to be carried by
higher energy particles (i.e. ions
\citep{ama03}). In principle the presence of ions might have
observable consequences in the TeV band, however even assuming the
remaining 20\% of the spin-down energy is all
carried by ions, their contribution to the gamma-rays emission
turns out to is negligible. On the other hand, if the return current
is made by a high energy lepton beam (Arons, in preparation), the radiative
consequences of this latter scenario have not been investigated,
however, again, they are not expected to give any appreciable contribution to gamma-rays

TeV emission is due to IC scattering by the pairs, for which the main
target is the nebular synchrotron emission in this case. We estimate 
a magnetic field $B \simeq 200\mu$G, slightly lower than
previous estimates \citep{aha04,hes08}. Overall the broad-band spectrum 
is very well reproduced, with the largest discrepancies limited to the {\it EGRET} 
data points around 10MeV. These points have large uncertainties and are likely affected
by calibration issues ( \citep{abd09}). Indeed recent {\it Fermi} data (\citealt{abdo10}) 
are consistent with our model curve.
\begin{figure}
\resizebox{\hsize}{!}{\includegraphics{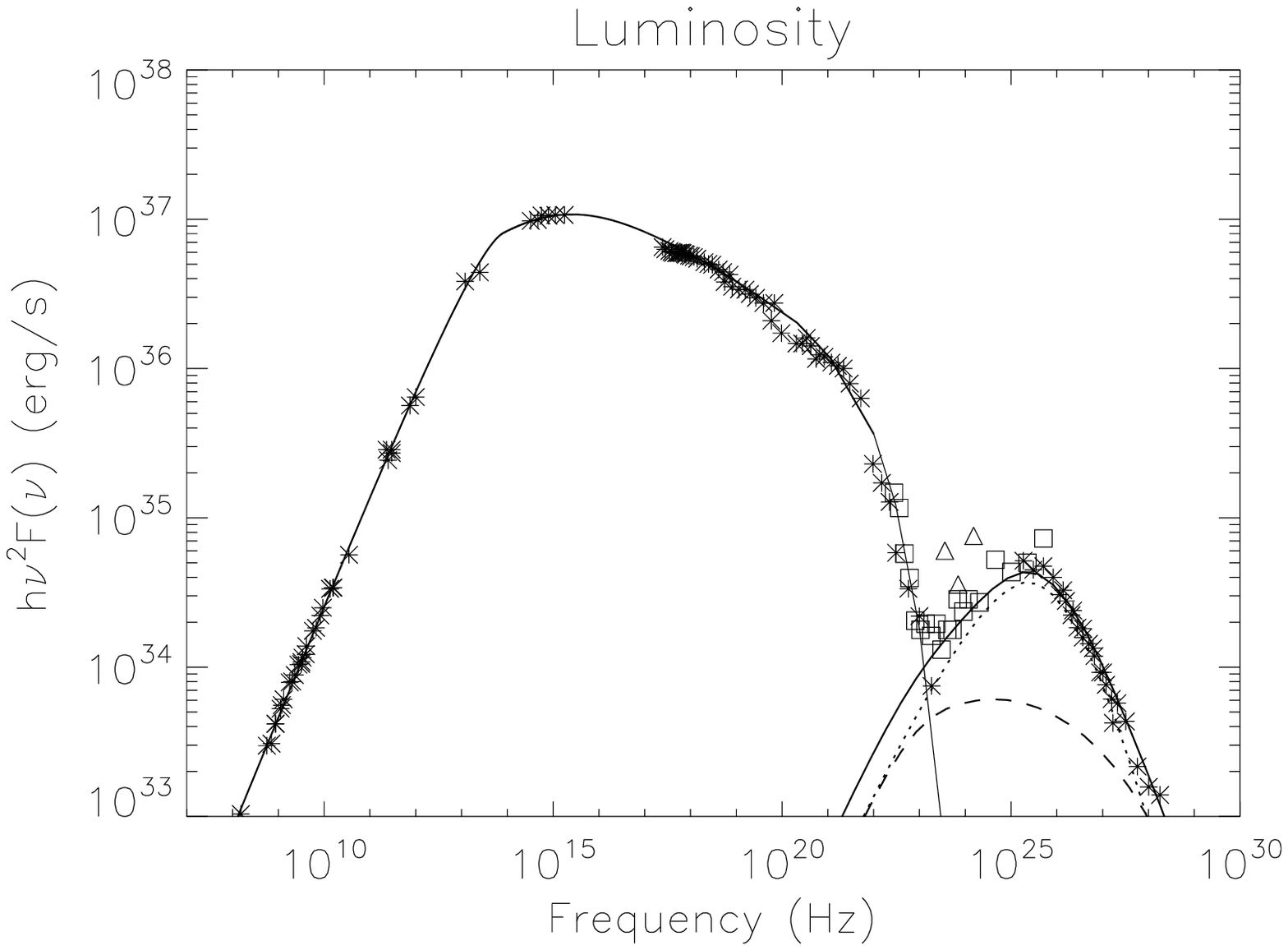}}
\caption{The Crab Nebula integrated emission spectrum. Data points are from
  \citet{bal70,baa72,mez86,ban02,ver93,hen92,kui01,aha04,alb08,abdo10}
  (triangles are {\it EGRET} points, squares are {\it Fermi} points). Solid
line is the total luminosity. Dashed line is the IC-CMB, dotted line
if the IC-SYN.}
\label{fig:crab}
\end{figure}

\citet{bal70} has shown that the radio power-law spectrum of the Crab Nebula
extends down to the ionospheric cut-off at $\sim 30$ Mhz. One can use this
piece of information to derive an upper limit on the minimum particle energy at injection. 
In order for the radio spectrum to extend to those energies as an
uninterrupted power-law, we need to assume $\epsilon_m/\epsilon_c <
7\times 10^{-4}$. From Eq.~\ref{eq:avgE} we then find for the wind Lorentz factor
$\gamma_w < 5\times 10^4$. This has to be compared with the typical Lorentz
factor of the particles at $\epsilon_c$, which is $7.4\times
10^5$, and with the minimum Lorentz factor of injected particles that
is $< 500$. The estimated value of $\gamma_w$ translates into a lower 
limit on the pair multiplicity of $\kappa \gtsim 10^6$. 

Our model allows us to compute {\it a posteriori} the energy radiation
losses. We do this in order to verify to what degree the adiabatic approximation
for the evolution of the PWN radius is correct.
For the Crab Nebula we find that about half of the total energy injected
into the nebula in the form of pairs has been lost via
synchrotron emission. However the dependence of the radius on the pulsar
luminosity $L(t)$ is in general weak (scales as $L(t)^{1/5}$ for
constant luminosity), so we expect at most modification of order 20\%
in the radial evolution, and maybe in the adiabatic losses. Such value
is well within the simplifications and the approximations of the
model. In order to properly take into account radiation losses, one needs 
to abandon any analytic solution for the dynamics, and solve the coupled
system of equations for the dynamics and the emission simultaneously.

\subsection{3C58}

3C58 is another example of a young PWN, which shares many similarities
with the Crab Nebula. It has a typical non-thermal spectrum extending
from Radio to X-rays \citep{sal89,tor00,gre92,sla04,sla08}. It shows clear 
evidence of energy injection from the central pulsar PSR J0205+64 in the 
form of a {\it jet-torus} structure \citep{sla02}.  Recent
SPITZER measurements of the PWN IR luminosity have showed clear 
evidence for a possible injection break, like in the case of Crab \citep{sla08}. 
At a typical distance of 3.2 kpc \citep{che05}, its size is about $5 \times 9$ pc, 
equivalent to a volume of $\sim 140$ pc$^3$ (corresponding to a spherical radius
$\sim 3.3$ pc). An association with the SN 1181 has been proposed
\citep{ste02}. However recent measurements of the expansion of the
filamentary structure surrounding the source \citep{bie06,rud07}, interpreted
through a simple model of the PWN-SNR system \citep{swa03,che05}, have ruled out
this possibility.

The pulsar present spin-down power is $L\simeq 2.7\times 10^{37}$ erg
s$^{-1}$, with a characteristic dipole age $t_c= 5390$ yr \citep{mur02}. Neither the
true age nor the PSR braking index are known, and this prevents any reliable 
determination of the initial spin-down power. However it is possible to obtain some
constraint from the information available on the expansion of the PWN and surrounding 
SNR. In particular, \citet{boc01} measured a mass in the filaments of $\sim 0.1
$M$_\odot$, and thermal emission corresponding to a typical shock velocity 
$\sim 340$ km s$^{-1}$. Assuming that the filamentary structure corresponds to 
the swept-up shell of ejecta (in analogy with the case of the Crab Nebula) then 
we can use our model for the system evolution and obtain an estimate of both
the system age and the PSR braking index depending on the density profile of
the ejecta.

In Table~\ref{tab:2}, we present a set of models derived for different
values of the unknown braking index (3 and 2.5), and for different density
profiles of the ejecta. In all the different models the swept-up mass and the 
nebular volume are consistent with extant measures. The shock speed varies
slightly, but is in all cases consistent with measurements, within 
the approximation of a spherical model. We again assume a canonical SN 
explosion energy of $10^{51}$ erg. As in the case of Crab the lack of any clearly
detected SNR shell prevents any reliable estimate of the ISM density and 
we simply assumed a fiducial value of $0.1$ cm$^{-3}$.
 
\begin{table}
\begin{minipage}{8cm}
\caption{Values of the parameters corresponding to different models for 3C58}
\label{tab:2}
\begin{center}
\begin{tabular}{l c c c c }
\hline
$\beta $ & $\alpha$ & $t$ (yr)& $M_{ej}$ (M$_\odot$) & $V_{SH}$ ( km s$^{-1}$)\\
\hline
2     & 0 & 2000 & 5.4 & 270 \\ 
2     & 1 & 2100 & 3.2 & 305 \\
2     & 2 & 2250 & 1.2 & 355 \\
2.33  & 0 & 2050 & 5.5 & 270 \\
2.33  & 1 & 2180 & 3.3 & 307 \\
2.33  & 2 & 2350 & 1.3 & 360 \\
\hline
\end{tabular}
\end{center}
\medskip
\end{minipage}
\end{table}

By looking at Table~\ref{tab:2}, one immediately realizes that the 
inferred age is not very sensitive to either the braking index or the 
assumed ejecta structure. The reason for this is that in all cases the age 
is much less than  the characteristic age, and the pulsar spin-down power 
in the past was not much larger than the present value. It is interesting to 
notice that in all models the shock speed tends to be higher for steeper ejecta,
favoring the case $r^{-2}$. However this would result in a very small mass of 
the ejecta, inconsistent both with estimates based on the observed filamentary
knots ($\sim 8$ M$_\odot$) \citep{rud07}, and with stellar evolutionary
models. 

It is our opinion that cases with a density profile which is either flat or $r^{-1}$
are more likely, even if they are associated with a lower shock
speed: this is because steeper ejecta lead to unreasonably small ejecta
masses. In all cases the age estimated for the system rules out any association
with SN 1181. However the filamentary structure, that we have
interpreted as mostly due to swept-up ejecta, is rather complex and with
various velocity components \citep{rud07}, so that it cannot be ruled out
that part of this filamentary network might be associated with the 1181 event.

\begin{figure}
\resizebox{\hsize}{!}{\includegraphics{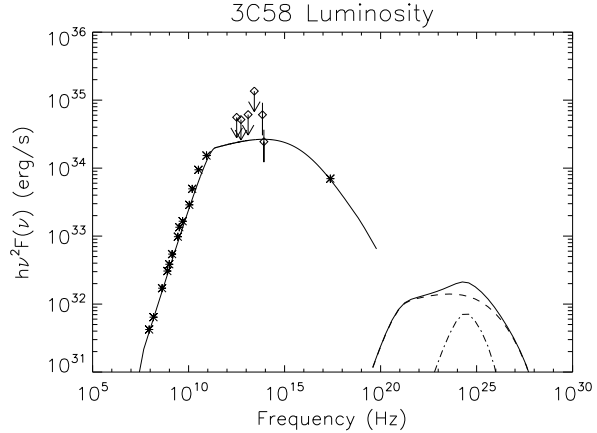}}
\caption{3C58. Data points are: radio from from
  \citet{sal89} \citet{gre92}, X-ray from \citet{sla04}, IR from \citet{sla08}
 (diamonds are IRAS and SPITZER points). Solid
line is the total luminosity. Dashed line is the IC-CMB, dotted line
is the IC-SYN.}
\label{fig:3c58}
\end{figure}

The question that naturally arises is whether spectral fitting allows to discriminate
between the models listed in Table~\ref{tab:2}. Unfortunately the answer is negative:
fits to the observations lead to similar results in terms of the required injection parameters
for all of these models. The reason is that all of them correspond to similar underlying ages, 
similar integrated spin-down power ({\it i.e.} total injected energy) and 
the same size (same adiabatic losses). The parameters reported in Table~\ref{tab:1} 
correspond to the case of ejecta with  $\alpha=1$ and braking
index  3. Spectral fitting is shown in Fig.~\ref{fig:3c58}. Very
similar values of the parameters are found also for a braking index of 2.5 ($\beta=2.33$).

If we compare the particle spectrum at injection with that of Crab, we find that
the spectral index is now flatter at low energies and steeper at high energies.
Again the required pair injection efficiency is very high $\sim 80$\% and the ratio 
$\mu_e \sim 8.3\times 10^{-5}$ is only a factor 2 lower than for Crab. The inferred 
magnetic field is $43  \mu$G, higher than other recent estimates \citep{sla08}. 
The ratio between the magnetic energy and the total energy is $\sim 0.5$, this is
quite high and leads to a value for $\eta_e+\eta_M >1$ which apparently violates 
the total energetics. The discrepancy is however $\sim 30\%$ and we consider it to be
within the approximations of our model (particularly critical in this respect are the 
simplifications related to the assumption of spherical geometry) and the uncertainties
of the data (especially the estimated distance, and hence
volume). Models with $\eta_e+\eta_M = 1$ always underpredict the radio
emission, independently of the value chosen for $\eta_M$.

Within the present model, we consider our estimates of $\eta_e$ and $\eta_M$ 
rather solid: attempts at fitting the data assuming a lower magnetization and a higher 
particles' content result in underproduction of the radio emission, while larger 
magnetization leads to the break frequency moving below $10^{11}$ Hz, in
contrast with observations. One final remark we want to make about $\eta_M$ is 
that the idea of a relatively high magnetization in this PWN is also supported by
its substantial elongation, created by hoop stresses of the toroidal
magnetic field \citep{begelman92}.

The existence of radio measurements down to 100 MHz allows us to derive also
in this case a lower limit on the pair multiplicity: we find that $\gamma_w < 3\times 10^4$ 
to be compared with the typical Lorentz factor of the particles at $\epsilon_c$, which is 
$1.2\times 10^5$, and with the minimum Lorentz factor of injected particles that
is $< 1600$. This implies $\kappa>5\times 10^5$. All the latter values are very close to
the corresponding estimates for Crab.

\subsection{MSH 15-52}

The system formed by the pulsar PSR B1509-58 and the SNR MSH 15-52 is 
striking for its relatively large size compared with its presumed age. 
Its distance is estimated to be $5.2\pm 1.4$ kpc \citep{gae99}. Assuming the
central value as a fiducial distance (which we will do in the following), the SNR radius 
turns out to be $\sim 83\times10^{18}$ cm. PSR B1509-58 is a young pulsar. 
Its current spin-down power is $1.8\times 10^{37}$ erg s$^{-1}$, with a 
characteristic dipole age of $1550$ yr \citep{kas94,liv06}. This is one of the very 
few pulsars for which the braking index is known: $\beta=2.087$. Unfortunately, 
our ignorance of the true age of the system still prevents knowledge of the initial 
spin-down power, $L_o$: this can only be determined as a function of the assumed 
age, $t$. A maximum possible age of the system can be estimated from the braking
index and the characteristic dipole age and the result is $t<1690$ yr. For
such an age, the size of the SNR implies a high $E_{SN}/M_{ej}$ ratio,
typical of SN Ib/c \citep{gae99,tsv02,maz00,iwa00,maz03,tan09}.
The size of the PWN is poorly constrained because of the high radio foreground 
from the SNR. Spectral modeling \citep{nak08,che05,dup95}, however, suggests 
that the cooling frequency lies just below the {\it CHANDRA} band (see also the present 
discussion), which is also in agreement with the size of the nebula at TeV energies 
being comparable with the {\it CHANDRA} size. If cooling starts being important at 
frequencies no lower than the X-rays, then one can estimate the size of the PWN 
from {\it CHANDRA} images: the result is a PWN radius of $\sim 17\times 10^{18}$ cm, 
for the assumed distance.

Knowing the braking index, and assuming an age ($t$) and
an ejecta density profile ($\alpha$), we can use our model to determine which
values of $E_{SN}$ and $M_{ej}$ return the assumed radius for both the
PWN and SNR. The only other parameter that enters this calculation is the
local density of the ISM. We assume $\rho_o=0.001$ cm$^-3$: a higher density 
would hint at even higher values of $E_{SN}/M_{ej}$ and in any case does not 
allow to match both radii; a lower value marginally affects the results because 
the system turns out to be in the ejecta dominated phase anyway.

If one accepts the hypothesis that B1509-58 was born in a SN Ib/c explosion,
then one can further constrain the model requiring that the mass of the ejecta 
is in the range $4-10 M_\odot$, and the kinetic energy released is in the range 
$5-20\times 10^{51}$ erg. In Tab.~\ref{table:3} we report the different age ranges. 
Flat ejecta tend to favor slightly younger systems. In spite of the relatively small 
differences between the ages reported in Tab.~\ref{table:3}, spectral modeling 
can prove a useful tool to discriminate between different scenarios for this source.
In the case of 3C58 the inferred ages for all models were much smaller than the 
characteristic dipole age, so that the injection properties of the PSR were almost 
the same, and it was not possible to discriminate using the SED. On the contrary,
for B1509, the inferred ages are close to the characteristic dipole age, and
different models correspond to different spin-down histories, and different
integrated injection energies. In such a situation the SED, and in particular
the radio data, can be used to rule out some of the models allowed by
the dynamics. 

Observations cover the radio GHz band \citep{gae99} (although the quality of
the data in this band is not very high), X-rays from {\it CHANDRA} to 
{\it INTEGRAL} \citep{gae02,del06,for06}, {\it Fermi} in the MeV-GeV range
\citep{abdo10b}, Cangaroo-II and {\it HESS} in the TeV \citep{nak08,aha05}. 
Only models with $L_o>10^{39}$ erg s$^{-1}$ can reproduce the
radio, while lower values of $L_o$ do not satisfy the energetic
requirements. Tab.~\ref{table:3} shows the limits imposed on the model
by SED fitting. Flat ejecta correspond to a very narrow range of possible
ages and a value of $\mu_e \sim 1.2\times10^{-6}$ smaller than what is
found both for the Crab Nebula and 3C58. Vice versa,  
ejecta with $\alpha=1$ correspond to spin-down properties and multiplicity
that are closer to what is found in Crab and 3C58. 
Compared with these two sources, we find a slightly harder high energy
spectrum at injection $\gamma_2 \sim 2.1-2.16$. However, this is 
consistent with the hard spectrum observed in the MeV-GeV range by 
{\it Fermi} \citep{abdo10b}. Indeed the photon index derived from
X-ray observations \citep{gae02,del06} of the inner region is $\sim
1.4-1.6$, and is much smaller than the previously used value $\sim 2.7$
which seems inconsistent with both observations and energetics
\citep{nak08}. Moreover the injection break is located between 5 and
50 GHZ, at a significantly lower frequency than what was assumed in
previous models \citep{nak08}.

Steeper ejecta give larger values of $\mu_e$, as reported in Tab.~\ref{table:3}.
All possible models require a high acceleration efficiency, $\eta_e \sim 0.7$ and a
strong magnetic field, $\eta_M \sim 0.53$ implying $B >
20\mu$G. Again, as in the case of 3C58, there is an energetic problem,
but as in the previous case, this system is strongly magnetized, as 
also suggested by the presence of a strong X-ray jet (see \citet{ldz04}
for a discussion of the dependence of the jet strength on the
magnetization).

The lowest frequency radio point allows us to constrain the
value of $\nu_e$, as it was done in Crab and 3C58. We find
that even for $\alpha=1$, the average wind Lorentz factor is
$\gamma_w < 10^4$, quite small compared with other systems.
This translates into a lower limit on multiplicity, that has to
be greater than $\sim 2-3\times 10^5$.  

\begin{table}
\begin{minipage}{8cm}
\caption{Different age rages for models for B1509-58}
\label{table:3}
\begin{center}
\begin{tabular}{l c c c c}
\hline
$\beta $ & $\alpha$ & $t$ (yr) Size lim. & $t$ (yr) SED lim. & $\mu_e$ \\
\hline
2.087     & 0 & 1350-1500 & 1470-1500 & $1.2\times10^{-6}$ \\ 
2.087     & 1 & 1570-1630 & 1570-1630 & $2-4\times10^{-6}$   \\
\hline
\end{tabular}
\end{center}
\medskip
\end{minipage}
\end{table}

\begin{figure}
\resizebox{\hsize}{!}{\includegraphics{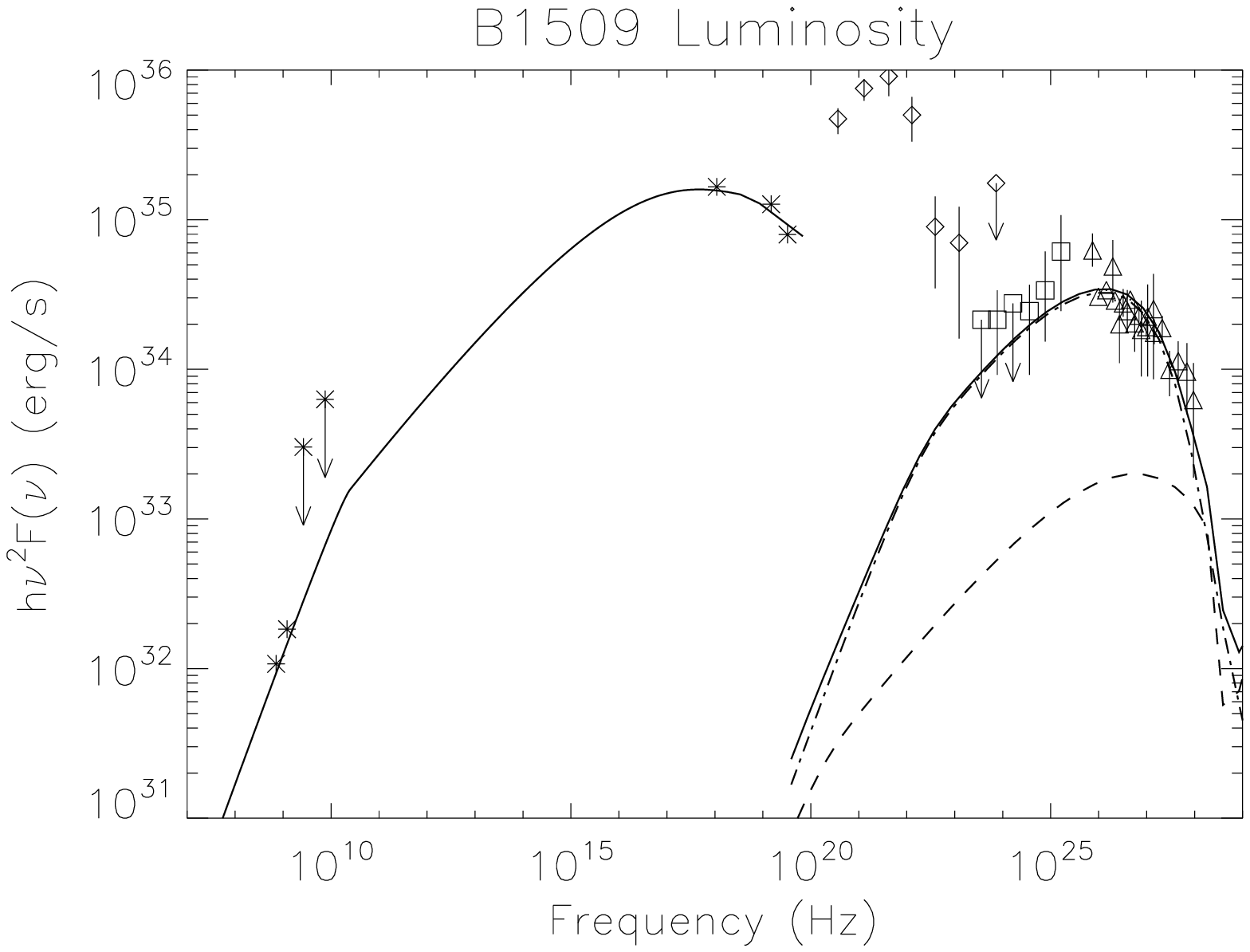}}
\caption{B1509-58; MSH 15-52. Data points are: radio from
  \citet{gae99}, X-ray from \citet{gae02} \citet{del06} \citet{for06}, $\gamma$-ray from
    \citet{nak08,aha05,kui99,abdo10b} (diamonds are {\it EGRET} points,
    squares are {\it Fermi} points, triangles are {\it COMPTEL}/{\it HESS} points). Solid
line is the total luminosity. Dashed line is the IC-CMB, dash-dotted line
is the IC on the enhanced local IR background at 400K. }
\label{fig:b1509}
\end{figure}

Fig.~\ref{fig:b1509} shows the spectrum derived from the model
presented in Tab.~\ref{tab:1}. From spectral fitting it is clear that
the Lorentz factor corresponding to $\epsilon_c$ has to be smaler
than $5\times 10^4$ ($\mu_e$ has to be smaller than $4\times 10^{-6}$), otherwise it is
not possible to reproduce the correct slope and intensity in the radio band (the
injection break shifts below 1GHz). As a result $\eta_M$ must be larger
than 0.3, and the magnetic field cannot be smaller than $15\mu$G. 
Only by measuring the PWN emission in the range $10^{10}-10^{12}$ Hz, 
it would be possible to discriminate among various models and put better
constraints on the multiplicity.

As it has already been noted \citep{aha05} it is not
possible to reproduce the observed gamma-ray emission in the TeV
range, either with IC on the CMB or the average galactic
background. The contribution from IC-SYN is negligible. 
Various models have been presented to account for such  discrepancy.

One suggestion is that of a lower value of the magnetic field in the nebula,
which would lead to infer a higher particle content, and consequently enhance 
the gamma over X-ray ratio \citep{aha05}. However, within this model this does
not seem to be a viable solution: in order to increase the particle content up to the 
value required to fit the gamma-rays as IC on the standard galactic background 
we would have to violate the energetic fixed by the spin-down history by a factor 
$\sim 5$. 

Another proposal is that of a possible contribution to the TeV flux of $\pi^o$ decay,
in the presence of relativistic hadrons \citep{nak08}. It was however immediately 
realized that this would require so much energy to be put in protons as to require
millisecond magnetar conditions at birth for the pulsar, which seems unlikely 
judging from the present PSR-SNR properties. In addition, the swept-up ejecta 
do not provide a sufficient target density of thermal protons.

The third possibility is that the local photon background, in particular
the IR, could be much higher than the average galactic value. Indeed
the SNR itself could be the origin of this excess \citep{nak08,dup95}. The
fit presented in Fig.~\ref{fig:b1509} assumes for the local IR
background a black body with a temperature $\sim 400$K suppressed by a
factor $3\times 10^{-7}$, which corresponds to an energy density about
5-7 times higher than the average galactic background. It is interesting, 
in this regard, the suggestion by \citet{helfand07} that the correlation between SNR
detected by {\it HESS} and HII regions might be due to enhancements in the
local photon background. 

\subsection{Kes 75}
The supernova remnant Kes 75 (G29.7-0.3) is a shell type remnant
which hosts at its center a young PWN powered by pulsar PSR
J1846-0258. This is a 324 ms pulsar, with a characteristic dipole age
of $723$ yr, a spin-down luminosity of $8.3\times 10^{36}$ erg
s$^{-1}$, and a measured value of the braking index $\beta=2.65$, which
provides an upper limit for the age of the system of $\sim 880$ yr \citep{liv06}.

The estimated distance of the system suffers from large
uncertainties. The first estimates pointed to a distance of 19-21 kpc
\citep{bec84}. However it was immediately realized that such a large value 
implied very peculiar conditions for the Supernova explosion \citep{mor07,hel03} 
and the ISM, together with an extremely
high synchrotron efficiency of the PWN. Recent studies have
revised the distance to a much smaller value, in the range 5-7 kpc
\citep{eis05,lea08}. We will assume a distance of 6 kpc in all our models.

Assuming a distance of 6 kpc, the radii of the PWN and of the SNR are,
respectively, $R=1.8$ and $R_{snr}=9.24$ ly. The velocity of the forward 
SN shock is $V_{SN}=3700\ {\rm km}\ {\rm s}^{-1}$ \citep{hel06} as 
estimated via the Si X-ray line. This places an upper limit on the age of the 
system $t<R_{SNR}/V_{SN} \sim 800$ yr. Various attempts have been made 
to measure the ejecta mass, and the SN energy; however the bulk of the SNR 
emission comes mainly from two regions of the SNR shell, and this tends to 
bias all estimates. In general, very high masses are inferred both for the swept-up 
ISM and for the SN ejecta \citep{bec84,mor07,lea08}, hard to reconcile with 
theoretical expectations for a SN Ib/c and standard evolutionary models. 
It must be stressed, however, that mass estimates are very sensitive to the 
assumed volume of the emission region, which is not well known. 

Proceeding as for previous cases, we will first attempt to constrain the possible 
SNR-PWN parameters through models for the dynamical evolution of the system. 
Given an assumed age $t$ and an ejecta profile $\alpha$ there is only one set of 
values of $E_{SN}$, $M_{ej}$ and $\rho_o$ that gives the correct size of both the PWN
and the SNR and the correct forward shock speed. Further constraints on the 
parameter space come from the properties of SN Ib/c which limit $M_{ej}$ in the
range $5M_\odot \simlt M_{ej} \simlt 16M_\odot$. Cases with $\alpha\ge 1$ are 
acceptable only for a system older than $700$ yr, and are in general unlikely
because they require a very low SN energy, $E_{SN}\simlt 0.3\times 10^{51}$ erg. 
Cases with flat ejecta are admissible for a larger range in age from 450 yr to 650 yr. 
The typical SN energy does not seem to depend on the age and is 
$E_{SN} \sim 2\times 10^{51}$ erg; younger systems require lower values of
$M_{ej}$, and larger values of $\rho_o$, as shown in Tab.~\ref{tab:kes75}. 
Cases with marginally steep ejecta, $\alpha=0.5$ suggest a larger age, a less
energetic explosion, $E_{SN}\sim 0.6\times 10^{51}$ erg, and lower
ISM density.

\begin{table}
\begin{minipage}{8cm}
\caption{Different age ranges for models for Kes75}
\label{tab:kes75}
\begin{center}
\begin{tabular}{l c c c}
\hline
$\alpha$ & $t$ (yr)  & $M_{ej}$ ($M_\odot$)  & $\rho_o$ (cm$^{-3}$) \\
\hline
0     & 450 & 4.8  & 4 \\ 
0     & 550 & 8.0  & 3 \\ 
0     & 650 & 16.4 & 2 \\ 
0.5   & 600 & 4.6  & 0.8 \\ 
0.5   & 700 & 10.7 & 0.5 \\ 
\hline
\end{tabular}
\end{center}
\medskip
\end{minipage}
\end{table}

The question arises again of whether it is possible to discriminate among the
different parameter sets for the dynamics based on the SED. Unfortunately this 
is not the case: as it was already realized earlier \citep{che04}, Kes 75 is a 
particle dominated system by large (more than 90\% of the pulsar spin-down power 
seems to be converted into accelerated particles, see below), so the differences 
in particle content due to the different energetics of the various models can be 
easily compensated by small changes in the magnetization to give the same
synchrotron emission.

However, fitting the model to  the multi-band emission spectrum is non-trivial. The radio 
emission spectrum implies a low energy particle spectral index at injection
$\gamma_1=1.7$ \citep{bec84,sal89,bok05}. The average photon index in the 
{\it CHANDRA} band is 1.9 while deep X-ray images have shown that the photon index 
in the vicinity of the pulsar is $\sim 1.8$ \citep{bla96,col02,hel03,ng08}. 

Evidence 
for a possible spectral break below 100 GHz was presented by \citet{bok05}, but
these data seem to be inconsistent with previous measurements \citep{sal89}
and a single power-law cannot be ruled out. 

Indeed in our model an injection break below 100GHz would require a very hard 
high energy injection, inconsistent with spectral information at X-ray frequencies. 
IR data do not provide good constraints \citep{mor07}. {\it INTEGRAL} data above 15 keV 
show a particularly hard spectrum, not fully consistent with an expected
cooling break in the {\it CHANDRA} band \citep{ter04}, but the pulsar might
contribute to the flux above 40 keV and be responsible for the excess
emission at high energies.

If the X-ray emission in the {\it CHANDRA} band corresponds to freshly injected high
energy particles, one must infer $\gamma_2 > 2.6$. However we found that
it is not possible to fit the overall SED when adopting such a soft high energy
injection. The reason is simple: given that the average spectral index is close 
to the one measured at the injection, the synchrotron break frequency must be 
above the {\it CHANDRA} band, but this implies a magnetic field $B\lessim 10\mu$G. Even
assuming $\eta_e=1$, with this small field all the models under-produce
the radio emission by at least a factor 5. We find that it is possible to fit
the synchrotron spectrum, including the average slope in the X-rays,
only if $2.2\le\gamma_2\le2.4$. 

\begin{figure}
\resizebox{\hsize}{!}{\includegraphics{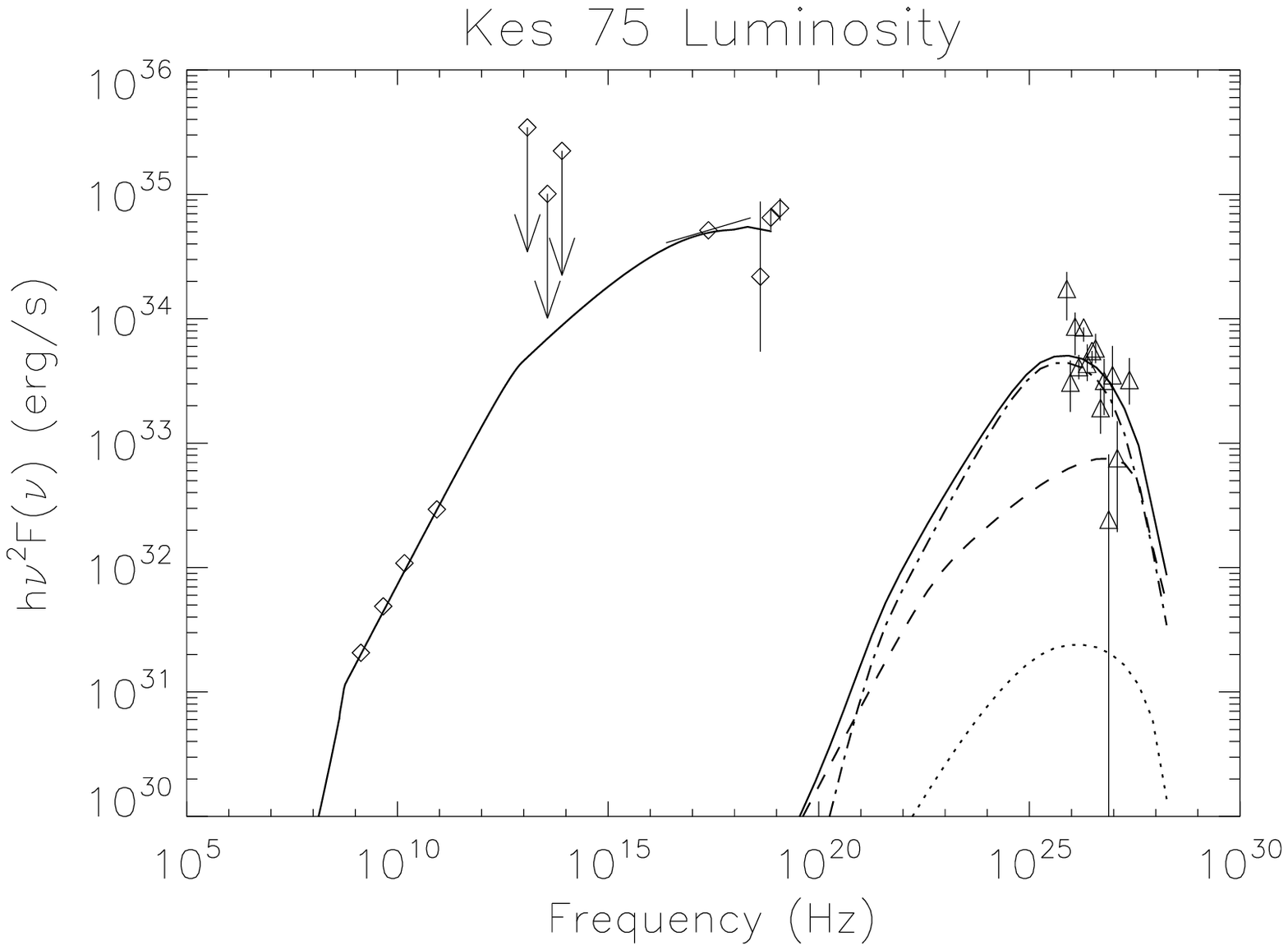}}
\caption{Kes75; SNR G 29.7-0.3. Data points are: radio and IR from
  \citet{bec84,sal89,bok05}, X-ray from \citet{col02,ter04},
  $\gamma$-ray from \citet{ter08,dja08}. Solid
line is the total luminosity. Dashed line is the IC-CMB, dash-dotted line
is the IC on the enhanced local IR background at 1000K. Dotted line is
the IC-SYN.}
\label{fig:kes75}
\end{figure}

As to the other parameters, we notice the following: the value of $\eta_M$ 
depends on age, with younger systems ($t\sim 450$ yr) requiring a field $B\sim 30\mu$G, 
and older ones ($t\sim 650$ yr) requiring $B\sim 20\mu$G, but it does not depend 
on $\gamma_2$; the value of $\mu_e$ depends on $\gamma_2$, ranging from 4000
for $\gamma_2=2.4$ to $4\times 10^4$ for $\gamma_2=2.2$, but it is independent of 
age. All the models require a very high injection efficiency $\eta_e
\sim 1$, to reproduce the {\it INTEGRAL} points. 
In Fig.~\ref{fig:kes75}, we show the spectrum derived according to the model in 
Tab.~\ref{tab:1}.
 
The Lorentz factor corresponding to $\epsilon_c$ ranges between
$2\times10^5$ and $1.5\times10^6$, very close to the values derived in
Crab. The one corresponding to the model of Fig.~\ref{fig:kes75} is
$7\times10^5$. The minimum Lorentz factor is lower than $\sim 5\times 10^4$,
and the average Lorentz factor is $\sim 7\times10^4$, with an associated
multiplicity that has to be greater than $10^5$.

Kes 75 has been detected in gamma rays by {\it HESS} \citep{ter08,dja08}. 
The particles responsible for TeV gamma-ray emission, in a purely leptonic 
model, are the ones emitting synchrotron radiation in the {\it CHANDRA} band. 
However the photon index in the TeV range is found to be $\sim 2.3$, far 
steeper than the average X-ray photon index of 1.9. Inverse Compton 
scattering on local synchrotron radiation gives a negligible contribution of TeV
gamma-rays. Despite this being a young system, PSR J1846-0258 has a low
spin-down luminosity, which results in a particle content about 2 orders of 
magnitude lower than in Crab and a correspondingly lower energy density
of the synchrotron photon field. Inverse Compton scattering on CMB photons 
is about 50 times stronger. However, even this contribution is about a factor 
of 10 below the observed flux. In addition, scattering on the CMB is expected
to occur in the Thompson regime and cannot account for the steepening of
the spectral index in gamma-rays. A suggestion that has been made is that 
this steeper spectrum hints at a much warmer radiation background, for which 
IC should take place in the Klein-Nishina regime. This requirement places the 
average black-body temperature of the seed background photons in the range 
1000-2000 K.  

With a black-body spectrum at 1000 K, suppressed by a factor $5\times10^{-9}$, 
it is possible to reproduce the correct gamma-ray luminosity. This contribution
corresponds to a local enhancement of the infrared background which is only a 
factor of a few above the galactic average. Interestingly, as in the case of B1509, 
also Kes 75 is surrounded by a bright SNR shell, that might be responsible for the
enhancement.

The question arises if the TeV gamma-rays can be explained by a
hadronic component, and in particular by $\pi^o$ decay. Given the very
low magnetization, inferred from the synchrotron spectrum, it can be
shown that only weak constraints can be put on $\eta_p$; in
particular one can obtain a good fit of the radio and {\it CHANDRA} X-ray data,
using $\eta_e\sim 0.5$ (which however undepredicts {\it INTEGRAL} data)
and a slightly higher nebular field, without
violating the energetics. However even by assuming half of the
spin-down energy goes into protons, in order to have a significant
contribution to the emission in the 10TeV range, at
least 10$M_\odot$ of target thermal protons are needed.  This is far in
excess of the swept-up ejecta mass $\sim 0.1M_\odot$, even if
consistent with the total ejecta mass (in this case one needs to assume
that protons escape from the PWN and interact with the SNR). The main
problem in this case is to properly reproduce the observed TeV
spectrum: our model predicts a peak at $10^{28}$ Hz, instead of
$10^{25}$ Hz as is observed, which is due to our choice of injection
energy for the protons, which is tied to $\gamma_w$. {\it Fermi} observations
should be able to distinguish between a leptonic and a hadronic model,
by constraining the emission below 1TeV. To properly investigate whether
$\pi^o$ decay is a viable possibility to explain the gamma-ray data, 
one would need a model for the diffusion of protons outside the nebula, which at 
the moment we do not have. The simplest possible estimate of the diffusion
time of protons outside the nebula, and in the ejecta, gives values that are about
an order of magnitude larger than the age of the nebula.

\subsection{Old Objects}

In this section we discuss two relatively old objects. The code has
been developed for the investigation of systems also beyond the free expansion
phase. However the late dynamics, especially if the pulsar kick velocity 
is important, can be quite complex. In addition old systems are usually 
very poorly constrained from the observational point of view. The age is 
generally known only as an order of magnitude estimate, and quite often 
either the central pulsar is not observed (as in the case of G327.1 and IC443), 
or the SNR is not observed. 

The following discussion highlights the difficulties that one has to face when 
trying to model sources for which both the quantity and the quality of the 
data are really poor. 

\subsubsection{W44}

The SNR W44 is known to contain an old PWN, associated with pulsar
PSR B1843+01. This is a 267 ms pulsar, with a characteristic dipole age
of $20,380$ yr, and a spin-down luminosity of $4.3\times 10^{35}$ erg
s$^{-1}$ \citep{wol91}. The value of the braking index is not known, while
the distance of the pulsar is estimated to be 3.1 kpc \citep{wol91} for a 
typical electron density $0.03$ cm$^{-3}$. 

The SNR W44 (3C 392) has an elongated shape with axes $25' \times 11'$.
Its distance is estimated from HI absorption to be 2.6 kpc, corresponding to a 
typical SNR radius of $11-13$ pc \citep{cox99}. The small discrepancy
between the estimated distance to the PSR and to the SNR is within the
uncertainties, and the association is considered a secure one.
This is a post-reverberation system,
for which pressure balance between the PWN and the SNR is important. The
central pressure is estimated from X-ray observations to be 
$\sim 1.4 \times 10^9$ dyne cm$^{-2}$ \citep{cox99}. 
Modeling the SNR \citep{cox99,she99} leads to the following estimates for
the relevant dynamical parameters: $E_{SN}= 10^{51}$ erg, $M_{ej}= 5 M_\odot$,
$\rho_o\sim 6$ cm$^{-3}$; for the expansion velocity of the SNR, HI emission
gives $v_{fs} \sim 150$ km s$^{-1}$ \citep{koo95}. A couple of remarks 
are here in order: first, all existing models assume the spin-down time of the pulsar
as the age of the nebula; second, the quoted value of $v_{fs}$ is inferred from 
a HI emitting ring structure which does not trace exactly the SNR, and 
estimates of the remnant speed can be as high as 330 km s$^{-1}$ \citep{koo95}. 

The PWN is observed in radio \citep{fra96,gia97} with a typical
luminosity of $200$ mJy. It has a distinctive cometary shape with the
pulsar located at the tip of a protruding finger of emission. The
nebula is thought to be a transition object from the spherical shape of
young systems to later bow-shock like morphologies \citep{swa04}. Within
this picture it is assumed that the PWN has already been crushed by the
reverse shock, and that the pulsar has been displaced by its proper motion 
with respect to the core of the radio emission. X-rays
have been detected both with {\it CHANDRA} and {\it XMM-Newton} from the head of the 
cometary nebula in the vicinity of the pulsar \citep{pet02,har06}, and the flux 
is measured to be $\sim 2.7\times 10^{-13}$ erg cm$^{-2}$ s$^{-1}$ in the 
2-10 keV band \citep{har06,pet02}. The radio emission is about twice as 
extended as the X-rays, suggesting effective cooling in the nebula, for 
which, however, there is no indication from the spatial behavior of the spectral 
index: this is found to be $\sim 2.2 \pm 0.3$ with no appreciable variations with 
the distance from the pulsar \citep{har06}. No IR or $\gamma$-ray emission is 
detected. At low radio frequencies the PWN is too weak to be detected against the 
SNR. 

The PWN is quite weak compared to the pulsar spin-down energy, and
appears also to be quite small in size, even if a correct determination of 
the volume is problematic, given the shape. At the distance of the SNR, a 
typical value for the volume is $\sim 0.5$ pc$^3$, corresponding to a radius
$\sim 1.5$ ly. It is interesting to notice that models of the PWN have
often assumed a much smaller age $< 5000$ yr \citep{pet02}, in order to fit the
observed spectrum, which clearly contradicts what has been used to model
the SNR. However such models are usually developed assuming that the
spectrum is Crab-like (in terms of the ratio between the radio and X-ray emission, 
and in terms of the location of the cooling break), but as we have shown through 
our modeling of young systems things can be different in different objects.  

The major problem in modeling this system, as was already recognized by \citet{pet02},
is the relatively low radio luminosity compared to the energy content of the 
nebula, and to the integrated pulsar spin-down power. This suggests that 
probably the age of the system is much lower. As a general rule, a younger age 
and a braking index less than 3 correspond to less injected energy. In particular,
we find that a braking index close to 2 and an age of about $10,000-15,000$ yr are
required to reproduce the observed radio emission. We also require that the 
average magnetic field in the nebula must be well below equipartition. Indeed we 
find that, in order to avoid overproducing the radio emission, $\eta_M$ has to 
be $<0.002$, corresponding to a nebular field below 10 $\mu$G, even if in principle 
this refers to the average magnetic field, while the value in the head can be higher. 

\begin{figure}
\resizebox{\hsize}{!}{\includegraphics{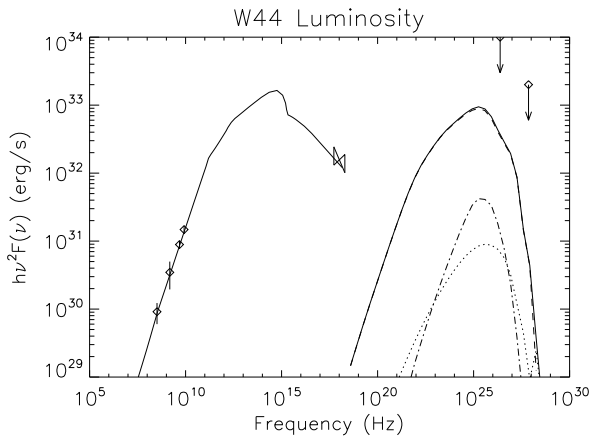}}
\caption{W44. Data points are: radio from \citet{fra96,gia97}, X-ray
from \citet{pet02,har06}, uppel limits on $\gamma$-ray from \citet{aha02,abdo09}. Solid
line is the total luminosity. Dashed line is the IC-CMB, dash-dotted line
is the IC on the local IR background (same as for Crab nebula). Dotted line is
the IC-SYN.}
\label{fig:w44}
\end{figure}

This small value gains some support
from the lack of spectral steepening with distance, which suggests that cooling is not 
important. Indeed the difference in size between the radio and X-ray nebula could be 
explained by the presence of a much stronger magnetic field in the head of the nebula 
than in the body, resulting in the suppression of X-ray emission in the latter. In 
Table~\ref{tab:old} we list the values of the parameters used to produce the curve in 
Fig.~\ref{fig:w44}. Values of the SNR parameters are not dissimilar from what has been 
used in the literature \citep{cox99,she99}. The value we find for the central pressure is 
$\sim 1.6 \times 10^9$, in agreement with X-ray observations, but the expansion speed 
is found to be higher, $v_{fs}\sim 300$ km s$^{-1}$, as a consequence of the younger 
assumed age. 

It is evident that the quantity and quality of existing data do 
not allow us to constrain the PWN model as much as it is
possible for younger systems. Moreover, the shape is far from
spherical, and geometrical effects connected with the existence of a bow
shock cannot be modeled correctly within out approach. 
However, by repeating the same analysis we did in the previous
section, we find that the parameter $\mu_e$ in the model is $8\times 10^{-3}$, which corresponds 
to a Lorentz factor at $\epsilon_c$ equal to $2\times 10^5$, while the the average wind 
Lorentz factor is found to be $\gamma_w\lessim 10^4$, corresponding to a multiplicity $\kappa \gtsim 10^5$.

One thing to notice is that the spectrum barely extends to X-ray frequencies. 
As pointed out above, this is due to the low value of the magnetization we adopt.
If the magnetization in the head is higher, as could be expected due to compression 
in the bow shock, this might enhance the X-ray emission. 

Another striking feature is the bump at around $10^{15}$ Hz: this is a typical feature 
of post-reverberation systems. During the reverberation phase, there is a typical 
energy at which synchrotron losses are balanced by adiabatic compression gains. 
Particles tend to accumulate at this energy, and this causes a bump in the spectrum, 
which survives at later times.
One final point of interest is the fact that for this system (and for relatively old systems 
in general) IC scattering on the CMB gives a TeV flux that is comparable with that in
the X-rays.

\begin{table*}
\begin{minipage}{17cm}
\caption{Model parameters for W44 and K2/3 corresponding to the curves
in figures \ref{fig:w44} and \ref{fig:kuk}. Units are the same as in Tab.~\ref{tab:1}.}
\label{tab:old}
\begin{center}
\begin{tabular}{l c c c c c c c c c c c}
\hline
Obj. & $t$  & $E_{SN}$  & $M_{ej}$   & $\rho_o$   & $\alpha$ &
$L_o$ & $\tau$ & $\beta$ & $\gamma_1$ & $\gamma_2$ & $\eta_e$ \\
\hline
W44 & 15000  & 1  & 5   & 3   & 0 &
0.018 & 22000 & 2.82 & 1.3 & 2.55 & 0.85\\
K2/3  & 8000  & 1   & 8   &  0.2 & 0 &
0.34 & 13750  & 2.66 & 1.2 & 2.82 & 0.95\\
\hline
\end{tabular}
\end{center}
\medskip
\end{minipage}
\end{table*}

\subsubsection{K2/3 Kookaburra}

The ``Kookaburra'' is a complex of compact and extended radio/X-ray
and $\gamma$-ray sources, that spans about one square degree along the
galactic plane. A large circular thermal shell with a broad wing in the
North-East and a narrower one in the South-West is revealed from
radio images. Diffuse X-ray emission and point like sources have been
detected by {\it ASCA} \citep{rob99,rob01}, {\it XMM-Newton} and {\it CHANDRA} \citep{ng05}. We
are here interested in the North-East wing where at radio and X-ray frequencies 
a nebula is found, hosting a young and energetic pulsar, PSR J1420-6048. 
The pulsar location is also coincident with TeV $\gamma$-ray emission
detected by {\it HESS} \citep{aha06}. 

PSR J1420-6048 is a 68.2 ms pulsar, with a characteristic dipole age of $13,050$ yr and
a spin-down luminosity of $10^{37}$ erg s$^{-1}$ \citep{dam01}. 
The braking index is not known: as for the case of W44, this implies that the true 
age could differ substantially from the characteristic dipole age. The distance of the 
pulsar is estimated to be 5.5$\pm 0.8$ kpc \citep{aha06}. 

Identifying the PWN associated with PSR J1420-6048 is, however, rather problematic. 
The wing-like structure hosting the pulsar is usually referred to as K2. In coincidence 
with the PSR an enhancement of radio emission, usually referred to as K3, is also observed. 
K2 has an extent of $\sim 15'\times10'$, a total flux at 20 cm of $\sim 1$ Jy, and a 
spectral index $0.2\pm 0.2$, while K3 has an extent of about $3'$, a total excess 
flux at 20 cm of $\sim 20$ mJy, and a spectral index $0.4\pm 0.5$ \citep{rob99}. No IR
detection has been reported.

X-rays have been detected by {\it ASCA} in the 2-10 keV band, with an extent
of about $7'$, an integrated flux of $4.8\times 10^{-12}$ erg cm$^{-2}$
s$^{-1}$, and a spectral index 1.4 \citep{rob01}. A much more compact ($0.5'$) but
extended source has been detected by {\it CHANDRA} around the pulsar, with a total 
flux, extrapolated to the 2-10 keV band, of 
$1.3\pm 0.14 \times 10^{-12}$ erg cm$^{-2}$ s$^{-1}$, and with a
spectral index $2.3\pm 0.9$ \citep{ng05}.

Gamma-rays were first detected by {\it EGRET} \citep{tho95}, and more recently by 
{\it HESS} \citep{aha06}. The total {\it HESS} luminosity in the 0.4-20 TeV band is 
$5.1\times 10^{34}$ erg s$^{-1}$, for the assumed distance, and the spectral 
index is found to be $\sim 2.2$. The 2$\sigma$ angular extension of the 
gamma-ray emission ($\sim 7'$) implies a nebular size $\sim 11$ pc.

One immediately realizes that from a dynamical point of view these data are 
somewhat confusing. In particular we think that the standard interpretation, 
according to which the small K3 region is the PWN, while the K2 region is the 
SNR \citet{rob01}, seems unlikely. Indeed, within the framework of our model,
this assumption leads to unphysical values for the supernova energy 
and ejecta mass and to overpredict the radio emission.

The larger extent of the {\it ASCA} source, compared to the {\it CHANDRA} one, 
goes against the expectation that the nebular size should shrink at higher frequencies 
because of synchrotron cooling. The difference in {\it ASCA} and {\it CHANDRA} 
spectral index might be due to uncertainties on the assumed value of N$_{H}$. 
The spectral index of the TeV emission (interpreted as IC-CMB) is consistent with 
the steeper {\it CHANDRA} spectrum, but not with the flat {\it ASCA} one. 

Even the radio data are not conclusive: if one assumes that the PWN also
contributes to the K2 emission, rather than just to K3, the radio emission changes by
about 2 orders of magnitude. The uncertainties in the age, size and
interpretation of the observed fluxes, in addition to the lack of a well defined
SNR, seriously hamper the modeling of the system. Indeed, as for W44, we 
will present here just a simple model fit to the data, without going into any 
detailed investigation of the parameter space. 

\begin{figure}
\resizebox{\hsize}{!}{\includegraphics{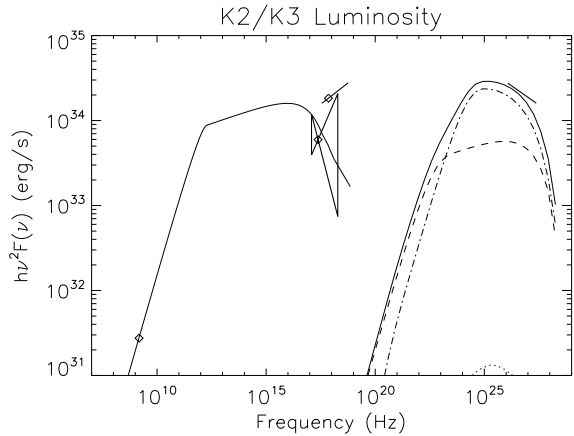}}
\caption{K3/2 Kookaburra. Data points are: radio from from
  \citet{rob99}, X-ray from \citet{rob01}, $\gamma$-ray from \citet{ng05,aha06}. Solid
line is the total luminosity. Dashed line is the IC-CMB, dash-dotted line
is the IC on the local IR background (assumed as a balck body at 200K). Dotted line is
the IC-SYN.}
\label{fig:kuk}
\end{figure}

Again, as in the previous case, the major difficulty in accounting for what is 
observed relates to the low efficiency both in X-rays and radio, together with 
the ratio $L_\gamma/L_X > 1$, which is suggestive
of a large nebular content of low-energy particles. As in the case of W44
this implies that a decent fit to the data can only be achieved if one
assumes a relatively young system, with a steep high energy spectrum (similar to
3C58) and a weak magnetization (comparable with that in Kes75). Fig.~\ref{fig:kuk} shows
the result of the model corresponding to the parameters in
Tab.~\ref{tab:old}.
The nebular radius is found to be $\sim 5$ pc in agreement with the
{\it HESS} size. The magnetic field is inferred to be $\sim 5\mu$G. It is
clear that it is not possible to reproduce simultaneously the {\it CHANDRA} 
and {\it ASCA} data. To get the steeper {\it CHANDRA} spectrum one needs 
to assume a cooling break around $10^{16}$ Hz. This is also necessary in 
order to reproduce the correct TeV spectrum. As already suggested by \citet{aha06}, 
in the TeV range the IC-CMB dominates over the IC on the standard galactic 
background. Indeed our model under-predicts the TeV emission by a factor 4-5. 
The local photon background must be higher if the observed TeV flux has to
be explained as IC emission: the TeV flux shown in Fig.~\ref{fig:kuk} was obtained
by assuming an IR background in the form of Black-Body at 200K,
with a photon density about a factor 4-5 higher than the  standard
galactic background. Given the 
complexity of the Kookaburra region, one cannot exclude such possibility.
  
The parameter $\mu_e$ in the model is found to be
$5\times 10^{-5}$,  which corresponds to a Lorentz factor at
$\epsilon_c$ equal to $2\times 10^5$, while the the average wind
Lorentz factor is $\gamma_w\lessim 10^4$, corresponding to a
multiplicity $\kappa \gtsim 10^5$.

\begin{table*}
\begin{minipage}{17cm}
\caption{Summary of the inferred lower limits on multiplicity.}
\label{tab:mult}
\begin{center}
\begin{tabular}{l c c c c c }
\hline
Crab & 3C58  & MSH 15-52  & Kes75   & W44   & K2/K3 \\
\hline
$10^6$ & $5\times 10^5$  & $3\times 10^5$  & $10^5$  & $10^5$   & $10^5$ \\
\hline
\end{tabular}
\end{center}
\medskip
\end{minipage}
\end{table*}

\section{Discussion \& Implications}
\label{sec:dis}
Let us briefly review here our findings, and what general conclusions can 
be drawn from the results of our attempt at modeling different objects,
both young and old.

\subsection{Summary of the results}

First of all, it is remarkable that our one-zone model, despite its simplifications, 
has proven able to account for the observed Spectral Energy Distributions in
all cases. This is already an interesting result, since the
assumption of efficient mixing that is at its base is likely a poor approximation at
high energies (in the X-ray band), where the flow pattern and the magnetic field 
structure in the inner nebula are expected to play a dominant role.  

The agreement between the model fit and the data is extremely good in the case 
of Crab, where the unknown parameters are reduced to a minimum (essentially the 
SNR properties). Our model seems to work rather well for young systems in general, 
but the quality of the data and the limited spectral coverage lead to larger uncertainties. 
In the case of strongly magnetized sources, like 3C58 and the nebula associated 
to PSR B1509, we find that in order to reproduce the data our model requires
an energy input in the PWN which exceeds the pulsar release by 20-30\%, judging 
from the current spin-down power. This conclusion does not seem to be affected 
by uncertainties in the distance to the sources: moving them closer to
(further from) us would lower (increase) the estimated brightness, but at the same 
time it would make them smaller and younger (larger and older), leaving the discrepancy 
between the radiated and accumulated energy almost unaltered. We think it likely that this
discrepancy is due to the simplifying assumptions of the one-zone model, which
become progressively more important at large magnetizations. In particular, as also
shown by recent multidimensional simulations \citep{vol08}, the radio and X-ray 
emitting particles might sample different magnetic fields. At the same time it is
worth keeping in mind that the estimate of the PSR spin-down power is based on the 
assumed canonical moment of inertia for neutron stars: $I=10^{45}$ g cm$^2$. 
A 20\% discrepancy in two systems is compatible with present uncertainties.
 
As one can easily see from Tab.~\ref{tab:1}, given the already high value of the 
quantity $\eta_e+\eta_M$, there is little energy ($\sim 20$\% at most) left to be
carried by any other higher energy component (ions or leptons in the
return current). The idea that a particle population with a larger Larmour radius
could be carrying most of the wind energy was made
attractive  by the following three main reasons: these 
particles could help solve the problem of 
particle acceleration at the pulsar wind termination shock; they could explain the
wisps variability in the Crab Nebula; finally, in the case of ions,
they could solve the discrepancy between the 
predictions of 1-d MHD models and the gamma-ray flux observed from Crab. 

As far as particle acceleration is concerned, at a relativistic pair shock, efficient
acceleration is shown to be possible only if the shock magnetization is extremely 
low \citep{spitkovsky08}. The 2-D MHD models of PWNe \citep{ldz05,camu09},
use a simple model of field reversal across the equator where a current
sheet occurs, such that  
the appropriate conditions for acceleration are thought to be realized only in a small sector of the 
pulsar wind termination shock, with no more than a few percent of the total wind 
energy flux flowing through it.

On the other hand, the presence of an energetically significant ion component in 
the pulsar wind is found to lead to efficient acceleration of the pairs \citep{hoshino92,amato06}.
This effect is the consequence of the large Larmor radius of the ions,
whose gyration introduces long wavelength turbulence through cyclotron
instability. The same result could come from the presence of high
energy leptons, accelerated as a consequence of runaway dynamics in
the equatorial current layer (Arons, in preparation).

As to the wisp variability, \citet{spitkovsky04} showed that this could be explained
as the result of compression waves associated with the high energy
current carriers' (ions, in their case)
 gyration in the shock
region: in order to reproduce the observed brightness contrast, again one would
need the current carriers (the ``beam'') to be energetically
significant: $\gamma_{beam} m_{beam}/\gamma_w m_\pm > 2 \kappa_\pm$. 
An alternate explanation of the wisp variability, which does without
the kinetic effects of the high energy particles beam, has been recently
shown by \citet{camu09}, who 
have proved that it can be recovered also within the framework 
of pure MHD, due to global instabilities of the termination shock.

One final consideration concerns the possible detection of signatures of
high energy protons. Modeling of the TeV emission from the Crab Nebula, based on the 
1D Kennel \& Coroniti flow \citep{atoyan96} was shown to underpredict the observed 
flux by a factor $\sim 5$, leading to suggest a possible contribution from the decay of 
neutral pions produced in nuclear collisions of relativistic protons. 
More recently, the presence of relativistic protons as the 
source of high-energy gamma-ray emission was suggested also in the Vela PWN 
\citep{horns06}. In the case of Vela, however, a later determination of the density of 
the thermal protons (that would serve as a target for nuclear collisions) resulted in too 
low a value and led to strongly question the initial claim \citep{lamassa08}. At the same 
time, \citet{vol08} showed that using a multidimensional model for the flow structure
in Crab, the estimated IC-TeV emission is easily overproduced, for the same parameters
that allow one to better fit the lower energy synchrotron emission. This suggests that
the results are indeed strongly model-dependent and that the discrepancies are
likely due to problems with the adopted MHD model. 

This is also the conclusion we reach in this work. We can reproduce the TeV emission 
from Crab, and find results in agreement with {\it Fermi} observations without any need for 
a proton contribution. In the case of both B1509 and 3C58, our pair+magnetic field 
energetic is already exceeding by 20\% the estimated PSR energy input, and fitting 
the gamma-rays through $\pi^0$ decay requires that protons alone carry far more energy 
than the nebula is currently estimated to store. Finally, in the case of Kes75, in order to 
ascribe the excess TeV emission to protons, one needs to assume that protons can 
effectively diffuse out of the nebula and experience the whole ejecta mass as a target. 

Our model seems to constrain with good accuracy the value of $\epsilon_c$. It is not 
possible to vary the value of $\epsilon_c$ by more than a factor of a few still satisfying
the overall energetics and reproducing the radio and X-ray data. At the same time, our 
model allows us to derive, for each object, an upper limit on the wind Lorentz factor and 
a lower limit on the pair multiplicity. The limits we put on $\gamma_w$ and 
$\kappa$ are strictly valid only if the wind has a unique Lorentz factor and the energy
scales $\epsilon_m$ and $\epsilon_c$ respect the assumed scaling with time, proportional
to the pulsar voltage. It is not easy to predict how a possible latitude dependence of the wind 
Lorentz factor, such as that included in axisymmetric models of PWNe ({\it e.g.} \citet{ldz04}), 
or a different dependence on time of the energy scales would affect our conclusions.

\subsection{Interpretation of spectral breaks}

Our study has reinforced a long standing puzzle: the electron (and
positron) accelerator at work in PWNe knows how to create a spectrum
convex in energy space, best represented by a broken power law.  As
was advanced by \citet{kennel84b}, the spectral steepening observed
between optical/soft X-ray energies and the harder ($\varepsilon > 10$
keV) spectrum is well understood as the effect of synchrotron energy
losses, with the pre-cooled spectrum a power law $N(E) \propto
E^{-\gamma_2}, \; 2.1 < \gamma_2 < 2.8$ from Table~\ref{parameters}.
The steepening between mid-infrared and optical, when observed,
requires energy space structure in the accelerator (an ``intrinsic
break''). The high energy spectrum can be attributed to
diffusive acceleration at the termination shock, at least qualitatively -
Fermi acceleration in relativistic very low sigma shocks in the test
particle limit yields a spectrum $N(E) \propto E^{-2.2}$
\citep{keshet05}. The radio spectrum requires a much
harder distribution of the lower energy particles, 
$N(E) \propto E^{-\gamma_1}, \; 1.2 < \gamma_1 <1.7$.
 
The standard model \citet{kennel84a, kennel84b}, developed to account
for the radiation from Crab at near infrared and shorter wavelengths, assigns
the conversion of the pulsar wind energy in the particle spectra
that emit the observed synchrotron radiation to diffusive Fermi
acceleration at the wind termination shock. Diffusive shock
acceleration always shows particle spectra with a Maxwellian
at low energy, plus a power law supra-thermal tail at high energy, with
the temperature of the Maxwellian set by the shock jump conditions
(except when the acceleration of the tail is very efficient). These
properties are well exhibited by Particle-in-Cell simulations of relativistic shocks in {\it
unmagnetized} $e^\pm$ plasmas (or, if magnetized, in upstream
quasi-parallel flow geometry \citep{spitkovsky08,  sironi09}).  Our
models identify the transition between the soft, high energy
spectrum and the hard, low energy spectrum as being at the energy
$\epsilon_c$, typically $10^5 - 10^6m_\pm c^2$ (see  Table~\ref{parameters}). 
In the \citep{kennel84b} model, this energy was
identified as the lower cutoff of the shock accelerated power law, and
that energy was identified as the ``temperature'' $\approx \gamma_w
m_\pm c^2$ of the flow downstream of the termination shock, thus
giving rise to the belief that pulsar winds have upstream flow Lorentz
factor $\gamma_w \sim 10^6$. 

The Kennel and Coroniti model and its descendants deliberately
neglected the radio emission from the Crab Nebula (and, by extension,
other PWNe.) The large (by number) population of particles with $E <
\epsilon_c$ suggests $\gamma_w$ to be much smaller, if the radio and
mid- to far-IR emitting particles come from the pulsar. That the
pulsars are the most likely source of the low energy particles in each
of the nebulae gains support from the existence of the radio ``wisp''
features in the Crab \citep{bie04} closely associated with the similar
time variable structures seen in optical and X-ray imaging, as well as
similar structure seen in 3C58 \citep{bie06} - those structures are
clearly coincident with the termination shock. Then,
in the regions of the flow populated by the large particle flux
feeding the low energy population, $\gamma_w \sim 10^4$ or smaller.

The low energy particle spectrum definitely is not a single
temperature relativistic Maxwellian, with $T \sim \epsilon_c$.  Since
the termination shock is not spherical, becoming more oblique in
higher rotational latitude regions \citep{komiss04, ldz04}, the
post-shock temperature declines with increasing latitude, suggesting
the low energy particles enter in the polar regions of the
outflow. However, replicating the observed spectrum as the envelope of
a sequence of Maxwellians requires a factor of $\sim 100$ variation in
temperature, which requires most of the mass flux being nearest the
rotation poles. There is no sign of pole to equator asymmetry in the radio
emission near the pulsars, except for the radio wisp features in the Crab, which are
components of the immediate post-shock flow, as judged from the optical and
harder photon emission. However we cannot rule out that mixing and
effects related to integration along the line of sight might prevent the detection of
the implied variations. 

However, let us assume in the following, that also the low-energy particle spectrum 
results from some acceleration process, rather than from the convolution of different
thermal distributions, and let us speculate on the nature of such a process.

Turbulence and associated Fermi II acceleration in and around the termination shock 
is one appealing possibility. The "wisp" motion, observed from the radio through X-rays
(see \citet{hes08} for a comprehensive review) has recently been interpreted as the
result of the strongly variable termination shock structure found in high resolution MHD
simulations of the Crab Nebula \citep{camu09}. The shock instability implied is a 
termination shock variant of the Standing Accretion Shock Instability, with outer scale variable
velocity $\delta v \sim v \sim 0.6c$ and length scale $\delta r \sim r\sim 0.5-1$ light years. 
The magnetized motions observed in the simulations (which do well in replicating the time 
variable spatial structure observed in the nebular ``wisps'') can act as an accelerator
through the Fermi II process \citep{kar62, staw08}, creating electron
and positron spectra $N(E) = N_0 (E/E_0)^{-s}$.  

In quasi-linear models of Fermi II acceleration in isotropic small amplitude Alfven
turbulence ({\it e.g.} \citet{staw08}), $s= \sqrt{9/4 + \epsilon} -1/2$, where 
$\epsilon = T_{accel}/T_{escape}$. These models include an analogue of scattering 
from large amplitude magnetic inhomogeneities (Fermi's original model), in the case 
where the wave energy spectrum scales in proportion to $k^{-2}$, with $k$ the
wavenumber. This case is germane to acceleration due to pair interaction with large 
scale moving magnetic fluctuations: the acceleration time scale for interaction with 
large scale moving ``eddies'' is $T_{accel} \approx \lambda_0/c (c/v_{eddy})^2 (B/\delta
B)^2 \sim$, where $\lambda_0$ is the outer scale of the turbulence, comparable to the
shock radius in our case \citep{camu09}.
The high downstream flow velocity suggests particles escape the turbulence zone through 
advective loss rather than diffusive escape (at energies much less than a PeV, microscopic
diffusion across B, even at the Bohm rate, is negligible compared to the advective losses). 
Then at flow speed $\sim c/3$, $T_{escape} \sim 3 R_{shock} /c$, leading to 
$\epsilon \sim (1/3) (B/ \delta B )^2 (\lambda_0 /R_{shock})$. Taking the parameters to 
be unity yields $\epsilon \sim 1/3$, in which case the accelerated spectrum is very
hard: $N(E) \propto E^{-1.3}$ which is similar to the spectrum
inferred for the radio emitting pairs in PWNe.  Because of the strong
radial mixing observed in the MHD models of PWNe, the
spectrum created in this turbulent acceleration zone will fill the
whole body of the synchrotron emitting nebulae, which provides a
natural explanation of the observed lack of gradients in the radio
spectral index of the Crab \citep{bie92}. Thus, we revive, in a
modern form, the suggestion of \citet{barnes73}, that the wisp motions
are responsible for a part of the particle acceleration required to
account for PWNe synchrotron emission.

Turning this idea into a full physical model may require extending the
MHD models to three dimensions. All models to date have been 2D,
axisymmetric, with exclusively toroidal magnetic field.  Particles
then cannot scatter radially, in order to sample adjacent moving
eddies, but are confined to toroidal flux tubes, which move in and out
in radius around average positions in the outflow (and inflow, at
higher latitudes) in the turbulent region.  It is possible that
magnetic pumping in these flux tubes might substitute for particle
scattering in quasi-isotropic turbulence as an acceleration mechanism
({\it e.g.}, \citet{mel69, kuij97}). However, polarization studies
already demonstrate that there are substantial poloidal magnetic
fields in PWNe inner regions, at least in the Crab.  The termination
shock instability identified by \citep{camu09} is not likely to be
restricted to 2D - perhaps the assumption of quasi-isotropic
turbulence, with forcing scale $\sim R_{shock}$, is the most natural
starting point for further pursuit of this idea. 

It is clear that a key point for the model to succeed is that the proposed
turbulently accelerated spectrum at low energy merges smoothly into
the shock accelerated spectrum at high energy. Since the macroscopic
turbulence is closely associated with the shock, a smooth merger is at
least thinkable. In this context the meaning of $\epsilon_c$ changes completely
with respect to the Kennel and Coroniti interpretation: this should not be identified with
the post-shock temperature (which according to the present model is about an
order of magnitude lower), but rather with the energy above which the low energy 
accelerator (Fermi II in the turbulent flow, in the suggestion made here) fades out and
relativistic DSA takes over.

\subsection{How Do We Move Forward?}

Multi-D high resolution MHD and PIC simulations of the termination
shock region separating the freely expanding wind from the
subsonically expanding nebulae will shed light on the possibly
turbulent flow, and on whether that turbulence can act as the low
energy accelerator discussed previously. A useful first
step would be to couple the diffusion-advection equation in energy space
to the MHD calculations. Observationally, infrared
observations of the young PWNe, using the Herschel space telescope
\citep{pilbratt09} as well as near infrared instruments on the ground
and in space, will greatly improve our understanding of the physics
behind the broken power laws in energy space inferred here for the
injected particle distributions.
These instruments will have more than sufficient angular resolution to
resolve the PWNe, thus emphasizing the need for multidimensional
models to quantitatively interpret the observations.

For all the objects studied, the plasma injection rates, here inferred
from the evolutionary models, not simply taken from the average over
the nebular lifetimes, exceed $10^5$ times the Goldreich-Julian
rate. Such high rates can be understood only as the result of pair
creation within the pulsars' magnetospheres - given the spectral
continuity in the SEDs, models which rely on the radio and far-IR
emitting particles, which dominate the injection rates, being
accelerated from the thermal plasma in and around the nebulae are less
likely than injection from the pulsars. Our results support and
extend a long held suspicion \citep{gallant02}, previously based only
on analysis of the Crab Nebula, that the particle loss rates from
young, high voltage pulsars are substantially in excess of the
inferences of particle {\it outflow} from all known magnetospheric
pair creation models. The answer may lie in intrinsic time dependence
of pair creation within pulsar polar caps ({\it e.g.}
\citet{timok09}), or in the so far unexplored outer magnetosphere
accelerators associated with the boundary layer return currents
separating the closed and open field lines \citep{aro09}. {\it Fermi} 
gamma ray observations are and will be useful in
constraining magnetospheric pair creation models.

\section*{Acknowledgments}
N.B. was in part supported by NASA through Hubble Fellowship grant
HST-HF-01193.01-A, awarded by the Space Telescope Science Institute,
which is operated by the Association of Universities for Research in
Astronomy, Inc., for NASA, under contract NAS 5-26555; and from NORDITA.
JA acknowledges the support of the US National Science Foundation (AST-0507813),
the NASA Astrophysics Theory Program (NNG06108G), and the taxpayers of California.

\vspace{-0.75cm}


\label{lastpage}

\end{document}